%% file: data_augmentation.tex
\documentclass[usenatbib,useAMS]{mn2e}
\usepackage[titletoc,toc]{appendix}
\usepackage{epsfig,psfig,graphicx,float,color}
\usepackage{subfigure}
\usepackage{mybibdefs}
\usepackage{multirow}
\usepackage{widetext}
\usepackage{url}
 \usepackage{color}
\usepackage{amsmath}

\newcommand\rr{\color{black}}

\usepackage{graphicx}
\usepackage{amssymb}
\usepackage{epstopdf}
\title[Data augmentation for machine learning redshifts]{Data augmentation for machine learning redshifts applied to SDSS galaxies}  
\author[Hoyle et al.]{Ben  Hoyle$^{1,2}$,Markus Michael Rau$^{1,4}$,Christopher Bonnett$^3$,Stella Seitz$^{1,4}$
\newauthor Jochen Weller$^{1,2,4}$ \\\\\\\\
$^1$Universitaets-Sternwarte, Fakultaet fuer Physik, Ludwig-Maximilians Universitaet Muenchen, Scheinerstr. 1, D-81679 Muenchen, Germany\\
$^2$Excellence Cluster Universe, Boltzmannstr. 2, D-85748 Garching, Germany\\
$^3$Institut de Fõsica d'Altes Energies, Universitat Autonoma de Barcelona, E-08193 Bellaterra, Spain\\
$^4$Max Planck Institute for Extraterrestrial Physics, Giessenbachstr. 1, D-85748 Garching, Germany\\
\\
{\tt E-mail: hoyleb@usm.uni-muenchen.de}
 }
  
\begin{document}
\date{Accepted ----. Received ----; in original form ----.}
\pagerange{\pageref{firstpage}--\pageref{lastpage}} \pubyear{2010}
\maketitle
\label{firstpage}
\begin{abstract}
We present analyses of data augmentation for machine learning redshift estimation. Data augmentation makes a training sample more closely resemble a test sample, if the two base samples differ, in order to improve measured statistics of the test sample. 
We perform two sets of analyses by selecting 800k (1.7M) SDSS DR8 (DR10) galaxies with spectroscopic redshifts. We construct a base training set by imposing an artificial $r$ band apparent magnitude cut to select only bright galaxies and then augment this base training set by using simulations and by applying the K-correct package to artificially place training set galaxies at a higher redshift.  

We  obtain redshift estimates for the remaining faint galaxy sample, which are not used during training. We find that data augmentation reduces the error on the recovered redshifts by 40\% in both sets of analyses, when compared to the difference in error between the ideal case and the non augmented case. The outlier fraction is also reduced by at least 10\% and up to 80\% using data augmentation.

We finally quantify how the recovered redshifts degrade as one probes to deeper magnitudes past the artificial  magnitude limit of the bright training sample. We find that at all apparent magnitudes explored, the use of data augmentation with tree based methods provide a  estimate of the galaxy redshift with {\rr  a low value of bias}, although the error on the recovered {\rr redshifts} increases as we probe to deeper magnitudes. These results have applications for surveys which have a spectroscopic training set which forms a biased sample of all photometric galaxies, for example  if the spectroscopic detection magnitude limit is shallower than the photometric limit.
\end{abstract}

\section{introduction}
\input{intro.tex}

\section{Data  and Augmented data}
\label{data}
\input{data.tex}
\section{Analysis and Results}
\label{results}
\input{results.tex}
\section{conclusions}
\label{conclusions}
\input{conclus.tex}

\section*{Acknowledgments} 
\label{ack}
{\rr The authors thank the anonymous referee for comments and suggestions which have improved the paper. }BH would like to thank Kerstin Paech for useful discussions. S.Seitz and M. M. Rau are supported by the Transregional Collaborative
Research Centre TRR 33 - The Dark Universe and the DFG
cluster of excellence ``Origin and Structure of the Universe''. CB:  Funding for this project was partially provided by the Spanish Ministerio de Economa y Com- petitividad (MINECO) under projects FPA2013-47986, and Centro de Excelencia Severo Ochoa SEV-2012-0234. Funding for the SDSS and SDSS-II has been provided by the Alfred
P. Sloan Foundation, the Participating Institutions, the
National Science Foundation, the U.S. Department of
Energy, the National Aeronautics and Space Administration,
the Japanese Monbukagakusho, the Max Planck
Society, and the Higher Education Funding Council for
England. The SDSS Web Site is http://www.sdss.org/. 

The Millennium Simulation databases used in this paper and the web application providing online access to them were constructed as part of the activities of the German Astrophysical Virtual Observatory.

\begin{appendices}
\section{MySQL queries}
We obtain observation and simulated data from the SDSS and the Millennium Simulation using the following MySQL queries.
\subsection{SDSS}
\label{sdss_q1}
To obtain SDSS data the following MySQL query is run in the DR8 and then separately in the DR10 schema:
\begin{verbatim}
SELECT s.specObjID, s.objid, s.ra,s.dec, 
s.z AS specz, s.zerr AS specz_err,
s.dered_u,s.dered_g,s.dered_g,s.dered_i,
s.dered_z,s.modelMagErr_u,s.modelMagErr_g,
s.modelMagErr_r,s.modelMagErr_i,s.modelMagErr_z,
s.type as specType, q.type as photpType
INTO mydb.specPhotoTable FROM SpecPhotoAll AS s 
JOIN photoObjAll AS q 
ON s.objid=q.objid AND q.cModelMag_u>0 
AND q.dered_g>0 AND q.dered_g>0 
AND q.dered_z>0 AND q.dered_i>0 
\end{verbatim}
\subsection{Millenium Simulation}
\label{ms_query}
The simulated galaxies are obtained using the following query in the Millennium Simulation MySQL interface
\begin{verbatim}
select galaxyId, ra, dec, z_app as spec_z, 
SDSS_u as u, SDSS_g as g, SDSS_r as r, 
SDSS_i as i, SDSS_z as z from 
Henriques2014a.cones.MRscPlanck1_SPM_0NUM  
where z_app >0.05 and z_app <1.0 
\end{verbatim}
This query is run by replacing the string SPM with M05 or BC03 to explore the different stellar population models, and the string NUM is replaced by the integers running from 1 to 3.

\end{appendices}
\bibliographystyle{mn2e}
\bibliography{photoz}

\end{document}

%% file: intro.tex
Photometric surveys can be maximally exploited for large scale structure analyses once galaxies have been identified and their positions on the sky and in redshift space have been measured. Measuring accurate spectroscopic redshifts is costly and time intensive, and is typically only performed for a small subsample of all galaxies. In particular the spectroscopic sample is often a biased sample of the full photometric galaxy catalog due to the limiting magnitude that a spectroscopic redshift for a galaxy can be measured, and the deeper limiting magnitude that a galaxy may be identified photometrically. This paper examines how the spectroscopic training set can be augmented (or complimented) to span an input feature space that more closely resembles that of the full photometric galaxy sample, to improve redshift estimates using machine learning.

Photometric redshifts can also be estimated by parametric techniques, for example from galaxy Spectral Energy Distribution (hereafter SED) templates. Some templates encode our knowledge of stellar population models which result in predictions for the evolution of galaxy magnitudes and colors. The parametric encoding of the complex stellar physics coupled with the uncertainty of the parameters of the stellar population models, combine to produce redshift estimates which are little better than many non-parametric techniques \citep[see e.g.,][for an overview of different techniques]{2010A&A...523A..31H,2013ApJ...775...93D}. Unlike non-parametric and machine learning techniques, the aforementioned template methods do not rely on training samples of galaxies, which must be assumed to be representative of the final sample of galaxies for which redshift estimates are required. Other Template methods are generated completely from, or in combination with, empirical data, however these templates both require tuning, and also rely upon representative training samples.

When an unbiased training sample is available, machine learning methods offer an alternative to template methods to estimate galaxy redshifts. The `machine architecture' determines how to best manipulate the photometric galaxy input properties (or `features') to produce a machine learning redshift. The machine attempts to learn the most effective manipulations to minimize the difference between the spectroscopic redshift and the machine learning redshift of the training sample.

The field of machine learning for photometric redshift analysis has been developing since \cite{2003LNCS.2859..226T} used artificial Neural Networks (aNNs). A plethora of machine learning architectures, including tree based methods, have been 
applied to the problem of point prediction redshift estimation \citep[see e.g.][for a further list and routine comparisons]{2014arXiv1406.4407S}, or to estimate the full redshift probability distribution function \citep[][]{2010ApJ...715..823G,tpz,2013arXiv1312.1287B,RauEtAllinPrep}. Machine learning architectures have also had success in other fields of astronomy 
such as galaxy morphology identification, and star \& quasar separation \citep[see for example][]{1997daa..conf...43L,2009arXiv0910.3770Y}.

One may combine machine learning techniques with template based methods, and with  knowledge of semi-analytic models, by augmenting the training sample with information drawn from templates or simulations. Previous work in this area was initiated by \cite{2004A&A...423..761V} who examined the use of synthetic SED templates \citep[e.g.,][]{1997A&A...326..950F} to augment the very small galaxy training samples available to them at the time in order to measure galaxy redshifts. Using data augmentation, the authors reduced the redshift error from 0.18 to 0.11 for 227 galaxies selected from HDF–N/S \citep[][]{2000A&A...359..489C,2000ApJ...538...29C} using aNNs as the machine learning architecture. They did not extend their analysis to the SDSS galaxies available at the time. {\rr More recently \cite{2009MNRAS.397..520W} use a hybrid empirical and template based $\chi^2$ approach using aNNs to improve the estimate redshifts for SDSS selected quasars, but do not extrapolate analysis outside of the training set.}

In this paper we extend this early analysis by using simulated galaxies drawn from the latest semi-analytic models with recent stellar population models, and by using standard template routines to augment a non-representative training sample of galaxies, to make it more closely resemble the `test' sample of galaxies for which redshift estimates are required. We also show how much one may rely upon this augmentation as the original training samples become more un-representative of the test sample.

{\rr
If the training sample covers the same input feature space as the test sample, but is biased with respect to number density, one may weight the training set galaxies to more closely resemble the test set galaxies. This method has been applied to the full probability distribution function of the redshift distribution using a K-Nearest Neighbour weighting scheme \citep[][]{2008MNRAS.390..118L}, and to individual galaxies \citep[][]{2009MNRAS.396.2379C}. This has also been applied by \cite{2014arXiv1406.4407S} using the `covariate shift' method. We do not use the covariate shift method, or other weighting schemes in this work, because the training and test sets are defined to be unrepresentative of each other, and instead rely on data augmentation to make the samples more similar. 
}

This paper is organized as follows: in \S\ref{data} we describe the data sample and the data augmentation process; we present the machine learning methodology, and the analysis and results using the augmented data in \S\ref{results};  discuss in \S\ref{diss}, and conclude  in \S\ref{conclusions}.

%% file: data.tex
In this study we use a mixture of observational data drawn from two SDSS data releases, combined with simulations and data which are augmented from the observational training data.

\subsection{Observational data set}
The observational data in this study are drawn from  SDSS Data Release 8 \citep[][]{2011ApJS..193...29A} and the SDSS Data Release 10 \citep[][]{2014ApJS..211...17A}. The SDSS I-III uses a 4 meter telescope at Apache Point Observatory in New Mexico and has CCD wide field photometry in 5 bands \citep[$u,g,r,i,z$][]{Gunn:2006tw,Smith:2002pca}, and an expansive spectroscopic follow up program \citep[][]{2011AJ....142...72E} covering $\pi$ {\rr steradians} of the northern sky. The SDSS collaboration has obtained  2 million galaxy spectra using dual fibre-fed spectrographs. An automated photometric pipeline performed object classification to a magnitude of $r\approx$22 and measures photometric properties of more than 100 million galaxies. The complete data sample, and many derived catalogs such as the photometric properties, are publicly available through the CasJobs server\footnote{skyserver.sdss3.org/CasJobs}.

The SDSS is well suited to the analyses presented in this paper due to the enormous number of photometrically selected galaxies with spectroscopic redshifts to use as training, cross-validation and test samples. We select galaxies from CasJobs with both spectroscopic redshifts and photometric properties using the query {\rr shown in \ref{sdss_q1}.}

{\rr The MySQL query} extracts model magnitudes from which we construct all possible colors combinations. We only examine model magnitudes and colors in this work so that we can trivially combine the observed data with simulations and augmented data. Recent work has shown an improvement to the machine learning redshift measurement by using more photometric properties as input features \citep{2015MNRAS.449.1275H}.

We further select galaxies that have the internal SDSS photometric galaxy classification $type=3$, have spectroscopic redshifts above 0.05, spectroscopic redshift errors less than 0.1, $r$ band magnitudes above $18$ and apparent magnitude errors below 0.4. This reduces the sample size to 802,590 galaxies for DR8 and 1,710,822 galaxies in DR10. The main differences between these data samples are that the DR10 sample probes to higher redshifts and deeper $r$ band magnitudes, see Fig. \ref{appmags}.

\subsection{Training, cross-validation and test samples}
In this paper we explore the effect of assigning redshift estimates for galaxies that fall outside of the feature space spanned by the original training set and cross-validation sets. This is to mimic the situation that the results of a system trained on a spectroscopic data set are applied to a (potentially deeper) photometric data set. Therefore we intentionally create a test set of galaxies that are fainter in the $r$ band than those galaxies used for training and cross-validation.

This is performed by applying an $r$ band apparent magnitude cut of 18.5 (20.5) to distinguish bright training and cross-validation sample galaxies from faint test sample galaxies for SDSS DR8 (DR10), which results in a faint galaxy sample of size 124,844 (304,455). We then perform data augmentation to modify and enhance the training and cross-validation samples so that they more closely resemble the test sample. This process is described in more detail below. 

We also ask the following question: How well would we have done, if we were to have had training and cross-validation data that are drawn from the same sample of galaxies as the test data? We refer to this benchmark analyses as the `ideal' case. To answer this question we build a separate training, cross-validation and test sample from the faint data sample, each of size one third of the full faint samples in both sets of analyses. We expect that the recovered redshift errors using the augmented data sets will be somewhere between the values using this ideal case, and the value obtained by having no training data that overlaps in feature space with the test data.

We follow standard machine learning methodology, such that the training sample is used to train the machine learning system for a given machine learning hyper-parameter set. The cross-validation sample is then used to select the best values for the hyper-parameters of the learned system. Once the best set of hyper-parameters has been decided upon, the test sample is used to measure the true ability of the learned machine to generalize to a new data set. Unless otherwise specified, the test sample in this analysis is always the ideal case test sample of faint galaxies, and is of size 33\% of the full faint sample in both sets of analyses. This choice of test sample size allows a fair comparison between the different combinations of data sets used for training and the benchmark ideal case.

\subsection{Data augmentation using simulations}
\label{semi-anal}
The bright data in both the DR8 and DR10 analyses are augmented using galaxies extracted from light cones \citep[][]{2012MNRAS.421.2904H} created from the Millennium Simulation \citep[][]{2005Natur.435..629S}. In detail we extract redshifts and SDSS estimated magnitudes of galaxies from the table \textbf {Henriques2014a} which uses the latest semi-analytic models \citep[see][for more details]{2014arXiv1410.0365H} and separately incorporate the stellar populations models of \cite{2003MNRAS.344.1000B} and \cite{Maraston:2004em}. We extract semi analytic galaxies from both sets of stellar populations models, using the {\rr query presented in \ref{ms_query}.}

{\rr The output of the query} produces a data set of size 4,859,249. We select subsamples of simulated galaxies by applying a lower limit $r$ band apparent magnitude selection of 18.1 (20.1) to mimic the faint galaxy sample of DR8 (DR10) for which we are interested in recovering redshift estimates. We note that this limit is slightly brighter than that of the faint galaxy test sample. This choice allows the machine learning framework to draw on a slightly brighter sample while training, which may aid the redshift assignment. We examine the effect of more carefully selecting training data \S\ref{knnsearch}. The final sample sizes are 346,116 (1,205,192) to be used in the DR8 (DR10) analysis of which we form a training set of size 66\% and a cross-validation set of size 33\%.

The semi analytic models have been tuned to match the abundance fractions of red and blue galaxies as a function of mass at redshift $z=0$. However the detailed color distribution of galaxies and their corresponding color evolution is described by the complex stellar population evolution physics combined with the predicted star formation and metallicity histories. We note that over the redshift range of interest here, the models have not been fine tuned to replicate the magnitudes and colors of observed galaxies and therefore provide an independent estimate of the galaxy distribution.

We do not require a test sample for the augmented data set because we are not interested in obtaining a redshift estimate for the augmented data at test time. We are using the augmented data sets to help in the training and cross-validation stages. We show that the addition of the augmented data in these stages will result in an improvement in the redshift estimate of the test set of faint galaxies. For brevity we refer to this simulated augmented data sample as `simulations' in what follows.

\subsection{Data augmentation using K-correct}
\label{kcorr-aug}
We use the SDSS K-correct package \citep[][]{2007AJ....133..734B} to augment the bright training sample. K-correct is able to estimate the apparent magnitudes that a galaxy of a given magnitude and redshift would have, if it were at a different redshift. 

K-correct performs a $\chi^2$ analysis by comparing the input galaxy magnitudes and redshift with different synthetic galaxy spectra to identify the best template to use as a base for which to approximate the evolution of the galaxy. The synthetic galaxy spectra are drawn from the \cite{2001ApJ...556..121K} and \cite{2003MNRAS.344.1000B}  stellar population models. In detail the templates explore a range of parameter space corresponding to different star formation histories with varying amounts of stellar metallicities and ages, and models for galactic dust extinction. The $\chi^2$ minimization is performed on weighted combinations of (up to five) SEDs, and the best fitting combinations are identified from sets of observation data, including earlier SDSS data releases. We refer the reader to \cite{2007AJ....133..734B} and {\tt howdy.physics.nyu.edu/index.php/Kcorrect} for more details.

We perform this data augmentation by first randomly selecting (with replacement) galaxies from the bright training sample. The values of the apparent magnitudes of the selected galaxies are Gaussian re-sampled with a scatter of $1\%$ of the measured SDSS magnitude errors. {\rr The K-correct package is deterministic and therefore this additional magnitude scatter allows the generation of similar, but distinct, augmented training examples.} We then assign a new redshift to the galaxy by sampling from a Gaussian of width 0.2 centred at the spectroscopic redshift. {\rr The choice of redshift resampling ensures that the resampled redshifts are not much larger than that measurable by the SDSS. We do not expect the results of this analysis to differ widely with other choices of redshift resampling}. We finally pass the spectroscopic redshift, the slightly re-sampled magnitudes and the new redshift to K-correct. K-correct computes an estimate of the apparent magnitudes that the galaxy would have if it were at the new redshift.

We apply the same apparent magnitude and redshift selection as in \S\ref{semi-anal}. The final sample sizes of K-correct augmented data are 435,172 (532,710) for the DR8 (DR10) analysis of which we again form a training set of size 66\% and a cross-validation set of size 33\%. We again note that we do not require a test sample for the augmented data. For brevity we refer to the K-corrected augmented data sample as `augmented' data in what follows.

\subsection{Visualizing the data samples}
\begin{figure*}
   \centering
 \includegraphics[scale=0.485, clip=true, trim=20 35 45 70]{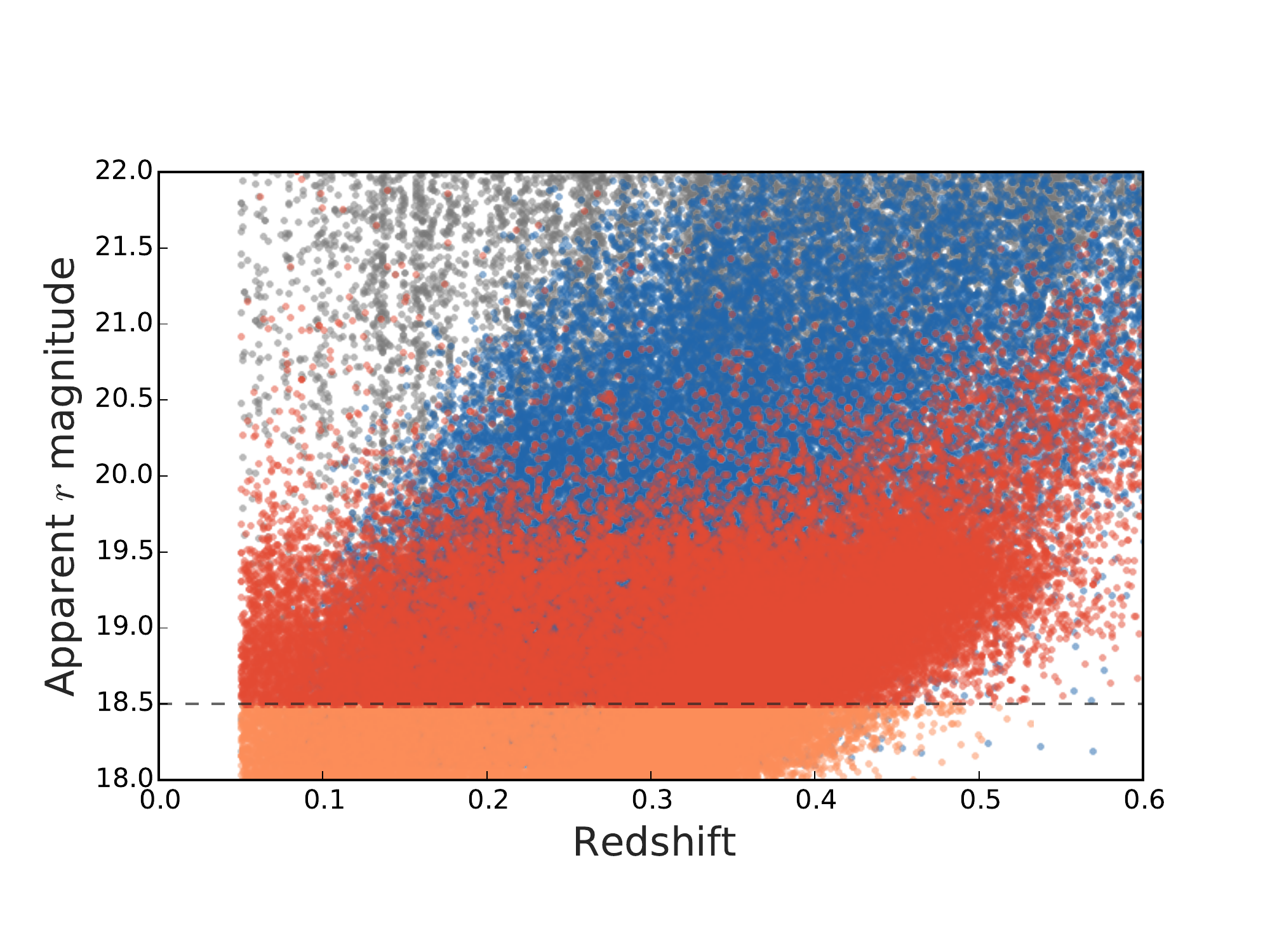} \includegraphics[scale=0.485, clip=true, trim=20 35 45 70]{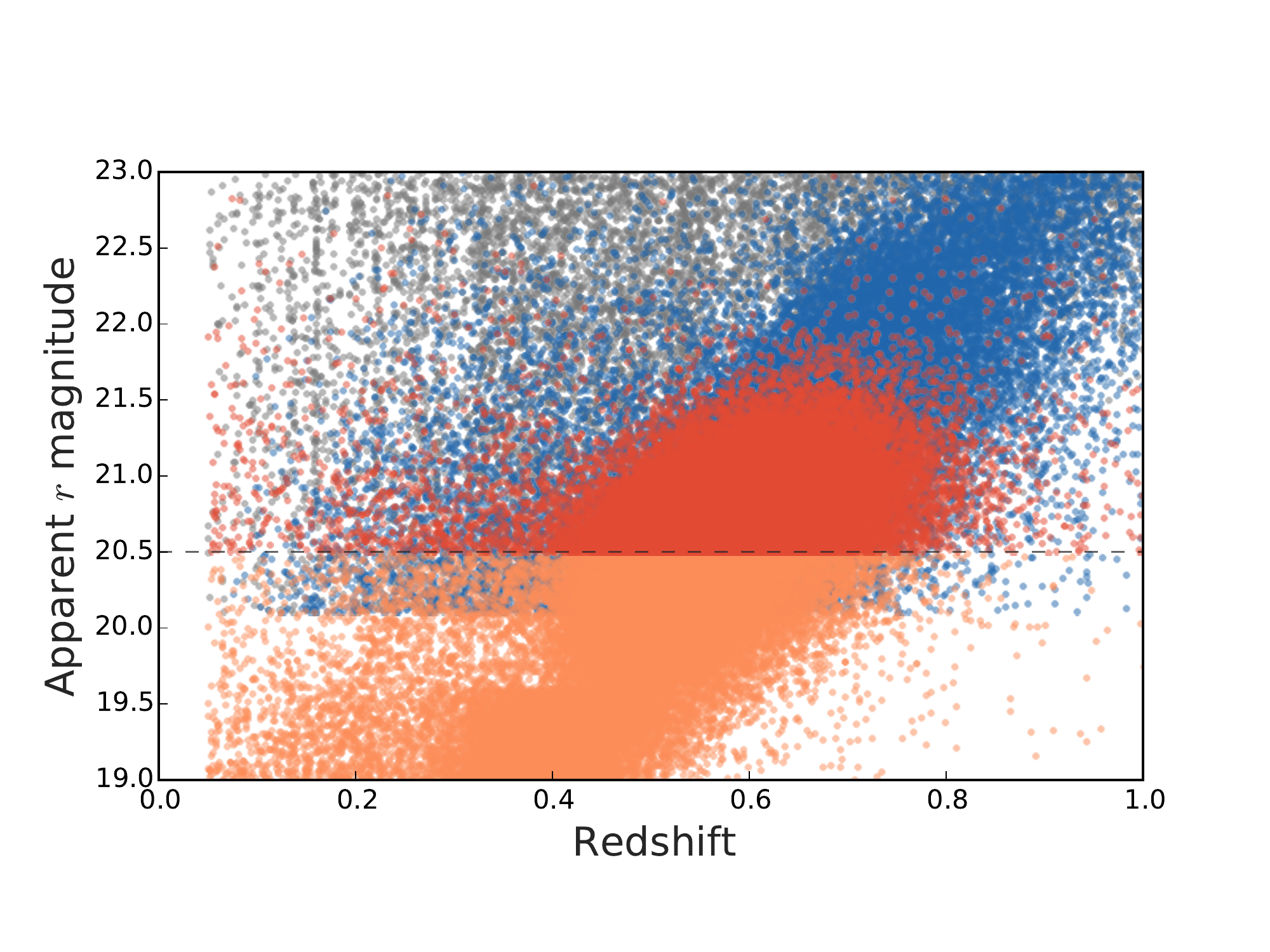} \\
 \includegraphics[scale=0.485, clip=true, trim=20 35 45 70]{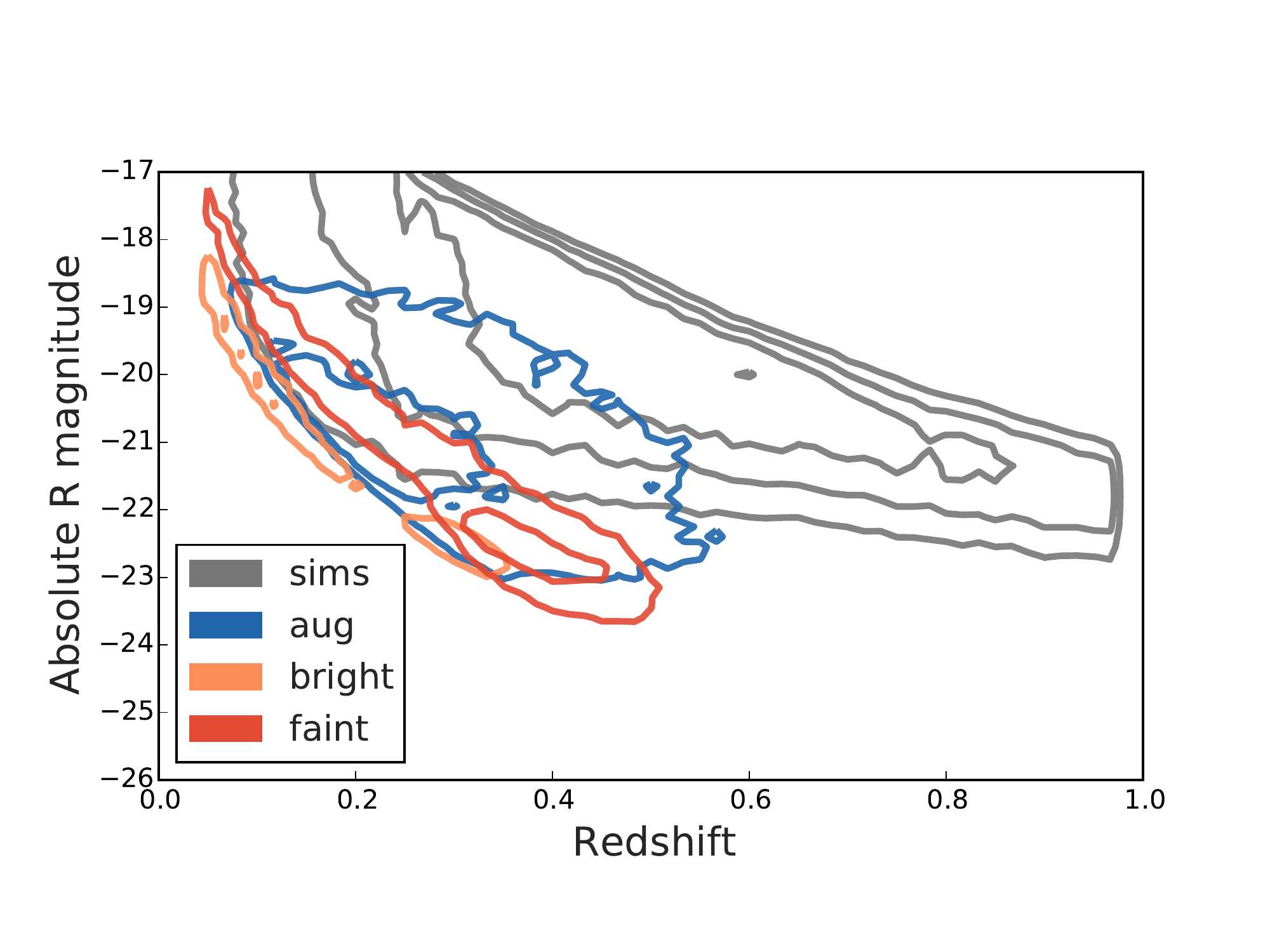} \includegraphics[scale=0.485, clip=true, trim=20 35 45 70]{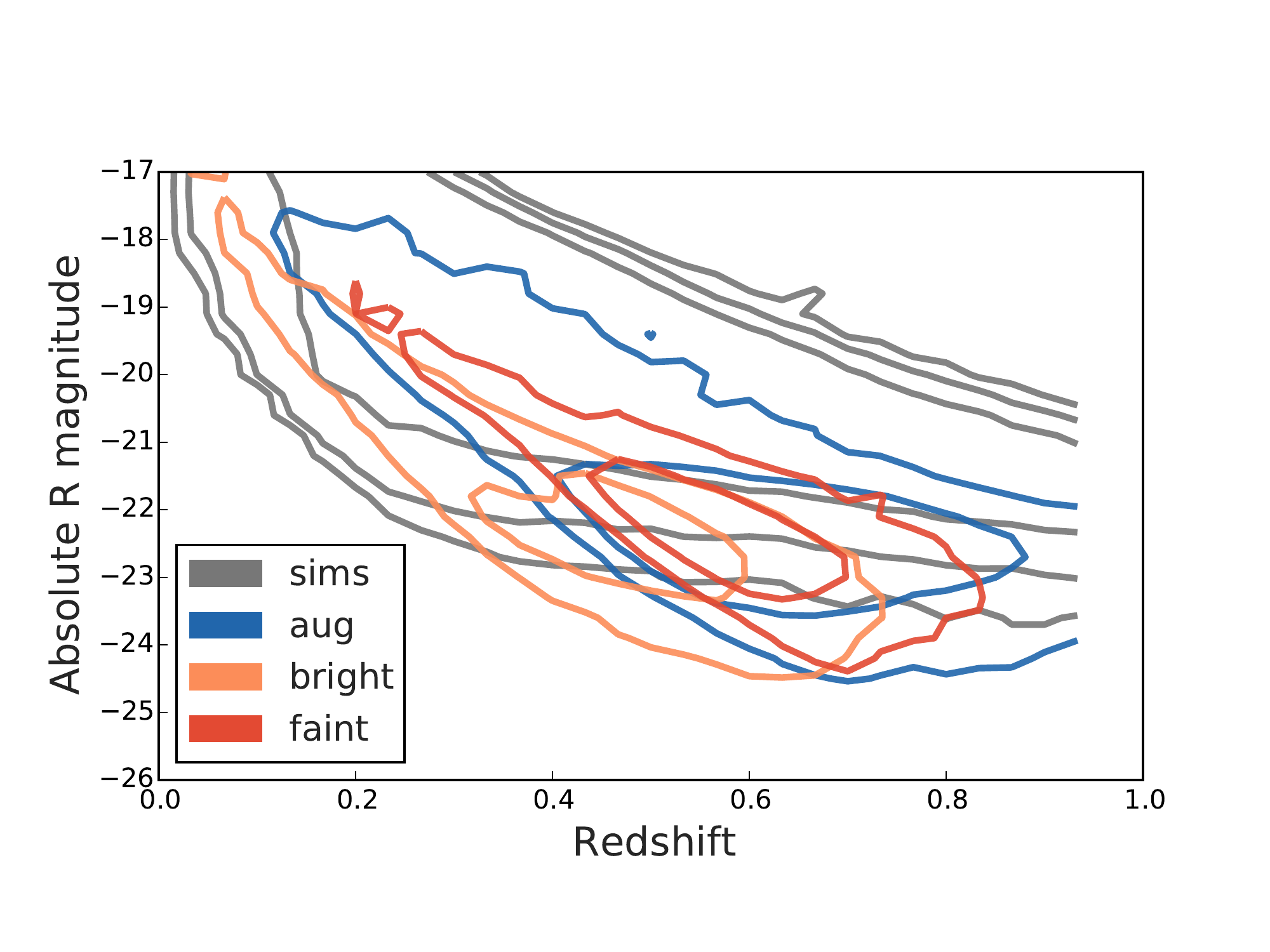} 
   \caption{ \label{appmags} The distribution of magnitude and redshift for a random selection of each of the SDSS DR8 (left-hand panels), and DR10 (right-hand panels) data samples. In the upper panels, the dashed line marks the separation between the faint test sample and the bright training sample. We modify the bright samples (in each distinct set of analyses) to generate the augmented data sample using SDSS K-correct. We further augmented the training set using the simulated galaxies produced using the latest semi analytic models. The top panel show a random selection of data, and the bottom panels show the absolute magnitudes density contours for each sample. In the legend `sims' denotes the augmented data using simulations, and `aug' corresponds to the K-corrected data augmentation of the bright galaxy sample.} 
\end{figure*}

In Fig.\ref{appmags} we present the $r$ band magnitude and redshift distributions of the different data samples. The left-hand (right-hand) panels correspond to the DR8 (DR10) analysis. We show the bright and faint observed galaxy samples, and the apparent magnitude limit separating these samples by the dashed line. We also show the simulated galaxy samples in grey and the augmented data in blue. For clarity we have only plotted a randomly selected subsample of the full data sets. The data points in the top panels show the redshift distribution against apparent magnitude, and bottom panels show density contours of the redshift distribution against absolute magnitude, as further determined by SDSS K-correct.

We note that both the K-correct augmented data and the simulated data occupy the same apparent magnitude space as the faint galaxy sample. It is due to this overlap that these augmented data sets are able to improve the redshift estimates of the faint test samples.

%% file: results.tex
We first briefly introduce the machine learning architecture employed in this work and then demonstrate how the addition of augmented data can improve the estimated machine learning (ML) redshifts for the DR8 and DR10 sets of analyses. We then show how the ML redshifts degrade as we push the analyses deeper past the apparent magnitude limits of the training sets. This is equivalent to extrapolating further into the unknown.

\subsection{Tree based methods}
The scikit-learn \citep[][]{scikit-learn} package written in Python has a very efficient implementation of decision trees for regression \citep[][]{ig}. The tree based machine learning architecture recursively partitions the input feature dimensions into an increasing number of bins. Each bin is chosen to minimize the scatter of the output feature, which for these purposes is the spectroscopic redshift. This results in data with very similar spectroscopic redshifts being within the same, or possibly nearby bins.

The power of tree based methods is enhanced by combining many trees. One technique to do this is called Adaptive Boosting or Adaboost \citep[][]{Freund1997119,Drucker:1997:IRU:645526.657132}, which adds trees sequentially to generate an ensemble of trees. In the following we will refer to this sample as a forest, but this term should not be confused with the term `random forest' which instead builds trees simultaneously. Adaboost weighs each new tree by its ability to predict redshifts correctly, and decides how trees are grown such that redshift estimates are improved for the data with poorly estimated redshifts.

We choose to vary the following sets of hyper-parameters; for a single decision tree: the number of data on each leaf node, for Adaboost: the loss function and the number of trees, while training: the training data set (e.g. the bright sample, or augmented data sample) and the size of the randomly selected training sample. For more details about combining trees with Adaboost we refer the reader to \cite{hastie01statisticallearning}\footnote{\url{statweb.stanford.edu/~tibs/ElemStatLearn}}.

We note that using an exponential loss function with Adaboost has been shown to behave poorly in the presence of classification noise \cite[see e.g.,][]{Dietterich:2000:ECT:350128.350131}. In this work we explore the three different loss functions available within the scikit-learn implementation, and we provide the final choice of hyper-parameters values in \S \ref{augemnet_all_data}.

\subsection{The effect of augmenting the data set}
\label{augemnet_all_data}

We perform separate analyses for the training and cross-validation bright data set and augmented data sets. We furthermore analyse different combinations of training and combined cross-validation samples, e.g. combinations of simulations and bright data, or simulations and K-correct augmented data.  We perform two independent sets of analyses, first on the SDSS DR8 catalog, with bright training and faint test samples defined by $r=18.5$ and then on the SDSS DR10 samples with bright training and faint test samples defined by $r=20.5$.

During the analyses we generate 200 distinct forests for each combination of training and cross-validation samples. For each forest we randomly choose the hyper-parameters, and draw a random sample of random size from the training sample for training. Once the forest has been trained we input the cross-validation sample to obtain a machine learning redshift $z$, and use this to determine the redshift scaled residuals, $\Delta_{z'}=(z-z_{spec})/(1+z_{spec})$. We calculate the value $\sigma_{68}$ from $\Delta_{z'}$ which is the value of the dispersion that encloses 68\% of the $\Delta_{z'}$, and is analogous to the standard deviation for Gaussian statistics.  We calculate the value $\sigma_{68}$ using the cross-validation set and select the forest with the smallest value as the winning forest.

At test time the faint test data is passed through the winning forest to obtain a redshift estimate. We next calculate $\Delta_{z'}$ and $\sigma_{68}$ as before, and additionally determine the outlier rate, defined as the percentage of data with $|\Delta_{z'}|>0.15$ \citep[following, e.g.,][]{2010A&A...523A..31H}, and the median $\mu$, of the distribution of $\Delta_z'$. We use test samples of size 33\% of the size of the faint galaxy sample size, corresponding to 41k for the faint SDSS DR8 galaxies and 100k for the faint SDSS DR10 galaxies. 

We present the results of the analysis performed on the SDSS DR8 sample in Table \ref{training_cvaldr8}. The top row shows the results of training and cross-validation on only the bright data. The quoted values are calculated on the test sample of faint galaxies, which we reiterate are never used in the training and cross-validation stages. This is equivalent to extrapolating the redshift estimate into area of input feature space which is unexplored by the training data. The extrapolation of analysis from a training set to an unrepresentative test set is poor machine learning etiquette, which is akin to extrapolating a result into the unknown. We include this analysis simply as a benchmark. 

The final row of Table \ref{training_cvaldr8} corresponds to a standard machine learning experiment. Here we train, cross-validate and test on random samples drawn solely from the faint galaxies. This is also included as a benchmark, and shows how well one might do in an ideal machine learning experiment, but this is not the main objective of this paper. All the other rows show combinations of training and cross-validations sets. The results of an identical analysis using SDSS DR10 sample is presented in Table \ref{training_cval}. We have also explored many other combinations such as training on the simulations, and cross-validating on the combined data, augmented data, and simulations, however the other combinations never perform substantially better or worse than the augmentation results listed in Tables \ref{training_cvaldr8}\&\ref{training_cval}.

\begin{table*}
\begin{center}
  \begin{tabular}{ | l | l | l | l | l |} 
Training & Cross-validation  & $\mu$ &$\sigma_{68}$ &Out. Rate\\ \hline
bright&bright& $0.0065$ & $0.0312$ & $2.8\%$\\  \hline
simulations&simulations& $0.0179$ & $0.0388$ & $5.65\%$\\
bright+augmented+simulations&bright& $0.0007$ & $0.028$ & $1.71\%$\\
bright+augmented+simulations&simulations& $0.0002$ & $0.028$ & $1.8\%$\\
bright+augmented &augmented& $-0.0001$ & $0.0273$ & $1.9\%$\\  
bright+augmented+simulations&bright+augmented+simulations& $0.0002$ & $0.0279$ & $1.77\%$\\
bright+augmented+simulations&augmented& $0.0008$ & $0.0281$ & $1.74\%$\\ \hline
faint&faint& $0.0$ & $0.024$ & $1.51\%$\\ \hline
  \end{tabular}
\caption{\label{training_cvaldr8} The values of the median $\mu$ and dispersion $\sigma_{68}$, and the outlier rate of the redshift scaled residuals calculated using the test set with 41k faint SDSS DR8 galaxies. The top row has no data augmentation, the last row presents the ideal case, and the rows in between use different augmented data sets for training and cross-validation.}
\end{center}
\end{table*}
\begin{table*}
\begin{center}
  \begin{tabular}{ | l | l | l | l | l |} 
Training & Cross-validation & $\mu$ &$\sigma_{68}$& Out. Rate\\ \hline
bright&bright& $0.0062$ & $0.0349$ & $1.76\%$\\ \hline
simulations&simulations& $0.011$ & $0.0554$ & $2.88\%$\\
bright+augmented+simulations&bright& $-0.0017$ & $0.0342$ & $1.69\%$\\
bright+augmented+simulations&simulations& $-0.0016$ & $0.0341$ & $1.74\%$\\
bright+augmented&augmented& $-0.0038$ & $0.0339$ & $1.68\%$\\
bright+augmented+simulations&bright+augmented+simulations & $-0.0025$ & $0.0338$ & $1.67\%$\\
bright+augmented+simulations&augmented& $-0.002$ & $0.0335$ & $1.73\%$\\ \hline
faint&faint& $-0.0026$ & $0.0315$ & $1.48\%$\\ \hline
  \end{tabular}
\caption{\label{training_cval} The same as Table \ref{training_cvaldr8}  for the analyses using 100k faint galaxies from SDSS DR10 as the test sample. }
\end{center}
\end{table*}

We see that most of the results from the data augmentation are between that of the best possible case (faint, faint) and the worse case (bright, bright), apart from the analyses using both the simulations training sample and the simulations cross-validation sample. However, while the values in this analysis are the poorest, we should note that the simulations assume nothing about the data in the redshift ranges of interest. They are using stellar population physics with observational anchors at z=0 and z=2. In itself this is still a remarkable result. We could have ignored all observed galaxy data, and K-correct augmented data and still obtain a redshift error of 0.039 for SDSS DR8 analysis (0.055 for the SDSS DR10), and an outlier rate $<5 \%$ ($<3\%$).

Examining the cases in both sets of analyses of training on combinations of bright data, augmented data, and simulations, and using the augmented data as the cross-validation set, we find that these values improve the redshift error by $10\%$ for SDSS DR8 ($4\%$ for SDSS DR10) compared to the bright data alone.  We find that the outlier rate in these cases improves by $40\%$  for the SDSS DR8 analysis (and a very modest $2\%$ for the SDSS DR10 analysis).

While these absolute values are of interest, it is perhaps more interesting to identify the level of improvement that we are able to achieve on the test sample with respect to the two benchmark cases. The benchmark cases correspond to having no augmented data (bright, bright) and the ideal ML case (faint, faint). We determine the relative ratio of improvement for each measured statistic with respect to these benchmarks using:
\begin{equation}
\textrm{I} = \frac{\textrm{(bright,bright)} - \textrm{(T,CV)} }{\textrm{(bright,bright)} - \textrm{(faint,faint)} }
\label{eq:RRI}
\end{equation}
where (T,CV) are the different training and cross-validation samples. For the case highlighted above (bright+augmented+simulations, augmented) we find that the relative ratio of improvement for $\sigma_{68}$ is 41\% for both the SDSS DR8 and DR10 analyses. The relative ratio of improvement for the outlier rate is 83\% for the SDSS DR8 sample and 11\% for the SDSS DR10 sample.

This means using data augmentation we are able to improve the redshift estimates from the worst case (bright samples) and recover up to 41\% (for $\sigma_{68}$) of the possible improvement that we may hope to achieve if we had the ideal case.

This analyses shows the power of using both augmented data, and simulated data, when estimating the redshift of galaxies in the cases when one has a training sample which is not representative of the test sample. For completeness we state the forest hyper-parameters of the best fit system using the data, simulation, augmented training and augmented cross-validation case. For DR8 this is {\tt  minleaf:24,numTrees:19, numTrainExamples:400320,loss:exponential} and for DR10 this is { \tt  minleaf:3,numTrees:56, numTrainExamples:659765,loss:square}.

We note that the smallest redshift errors in both sets of analyses, occur when the training and cross-validation samples are truly representative of the test sample, i.e. in the ideal case. We do not find that the addition of augmented data to the ideal case improves the recovered redshift estimates.

\begin{figure*}
   \centering
   \includegraphics[scale=0.46, clip=true, trim=0 0 40 35]{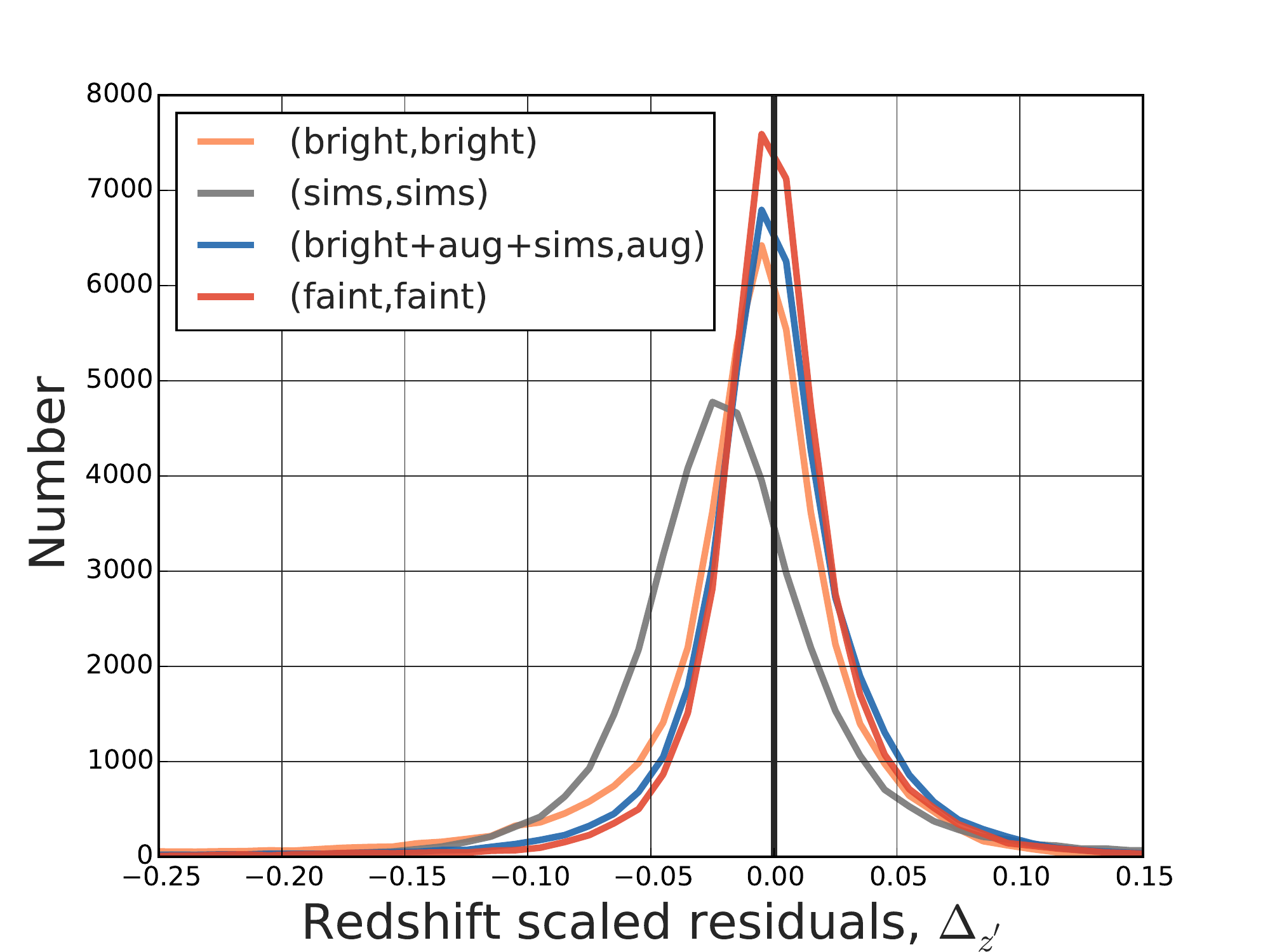}
 \includegraphics[scale=0.46, clip=true, trim=0 0 40 35]{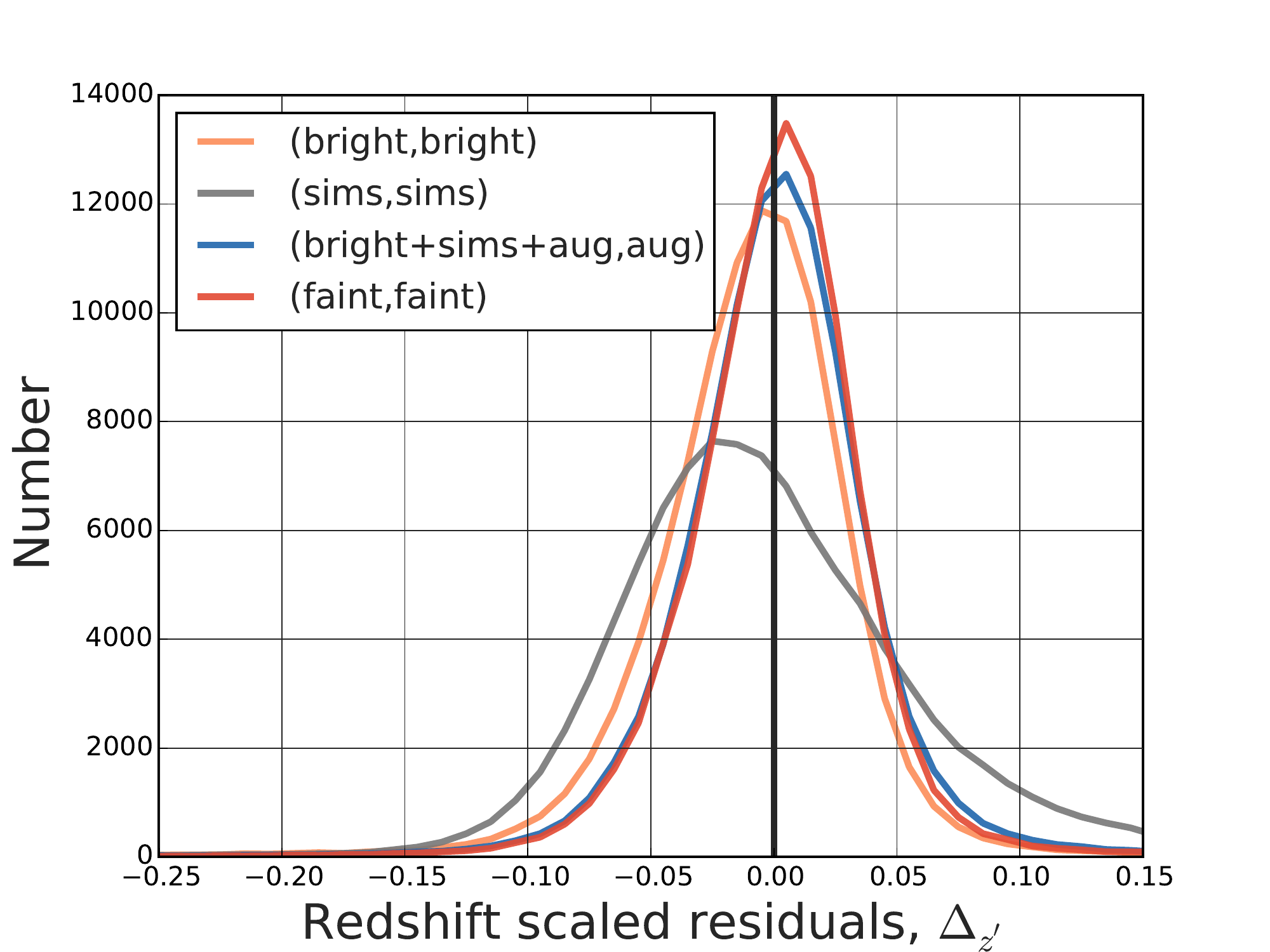}
   \caption{ \label{redshiftPlots} The distribution of test sample redshift residuals scaled by $1/(1+z)$ for different combinations of training and cross-validation samples used in the machine learning process (see legend). In the legend `sims' denote the simulations augmented data, and `aug' corresponds to the K-corrected data augmentation of the bright galaxy sample. For clarity only a few combinations are shown. The left panel corresponds to the DR8 analysis, and the right panel corresponds to DR10 analysis. The lines show the results using the test sample which presents an unbiased estimate of the true redshift error because it has not been used in the training or cross-validation processes. } 
\end{figure*}

In Fig. \ref{redshiftPlots} we show the distribution of $\Delta_{z'}$ for a selection of the analyses listed in Tables \ref{training_cvaldr8}\&\ref{training_cval}. We note that all distributions are both more peaked and have longer tails than a Gaussian distribution, which motivates our choice of $\sigma_{68}$. Also note the offset of the peak in the distribution when using the simulations as training and cross-validation sets, this can also been seen in the results Tables \ref{training_cvaldr8}\&\ref{training_cval}.

\begin{figure*}
 \includegraphics[scale=0.475,clip=true,trim=5 40 50 35 ]{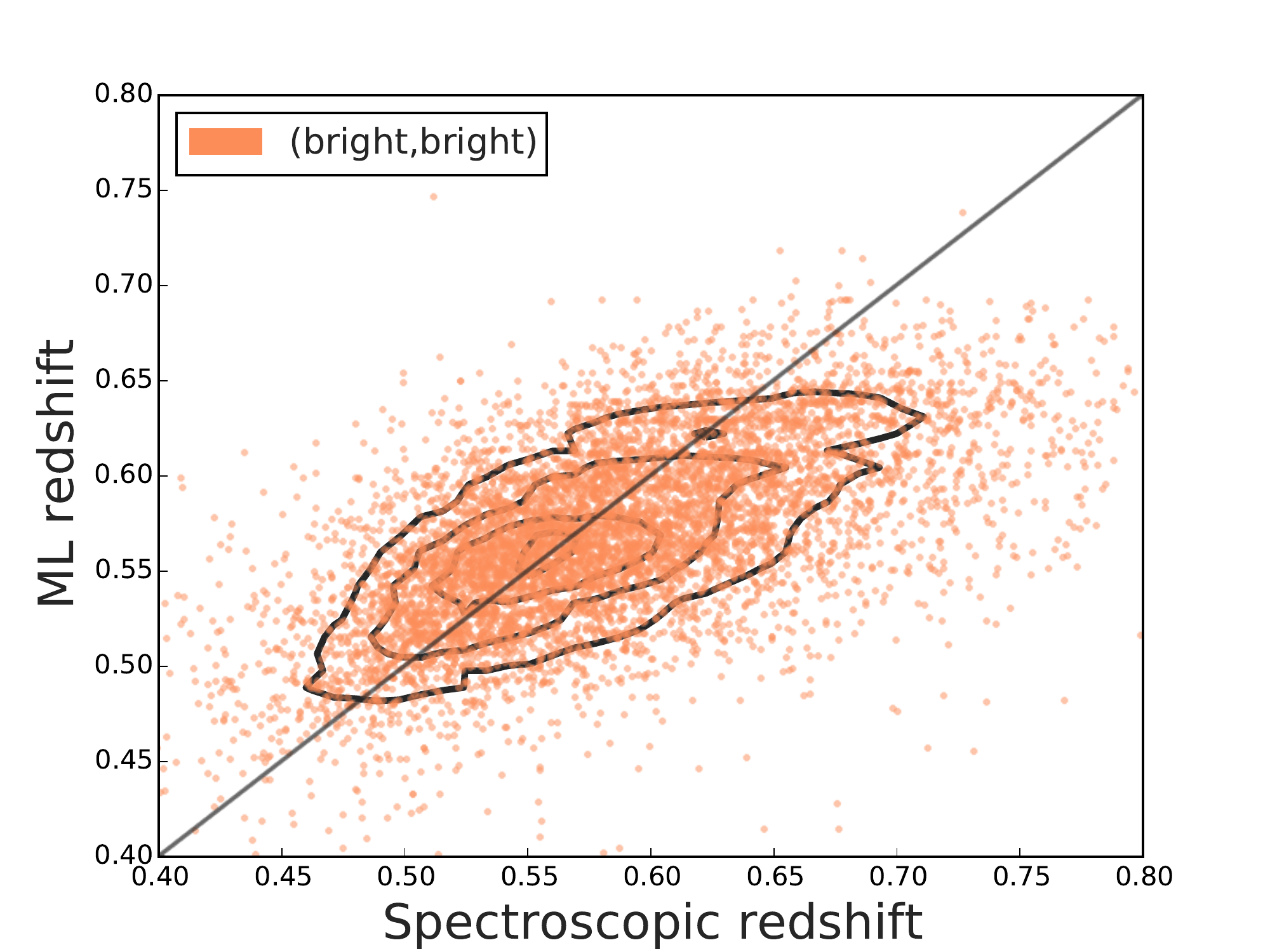}
  \includegraphics[scale=0.475,clip=true,trim=70 40 50 35 ]{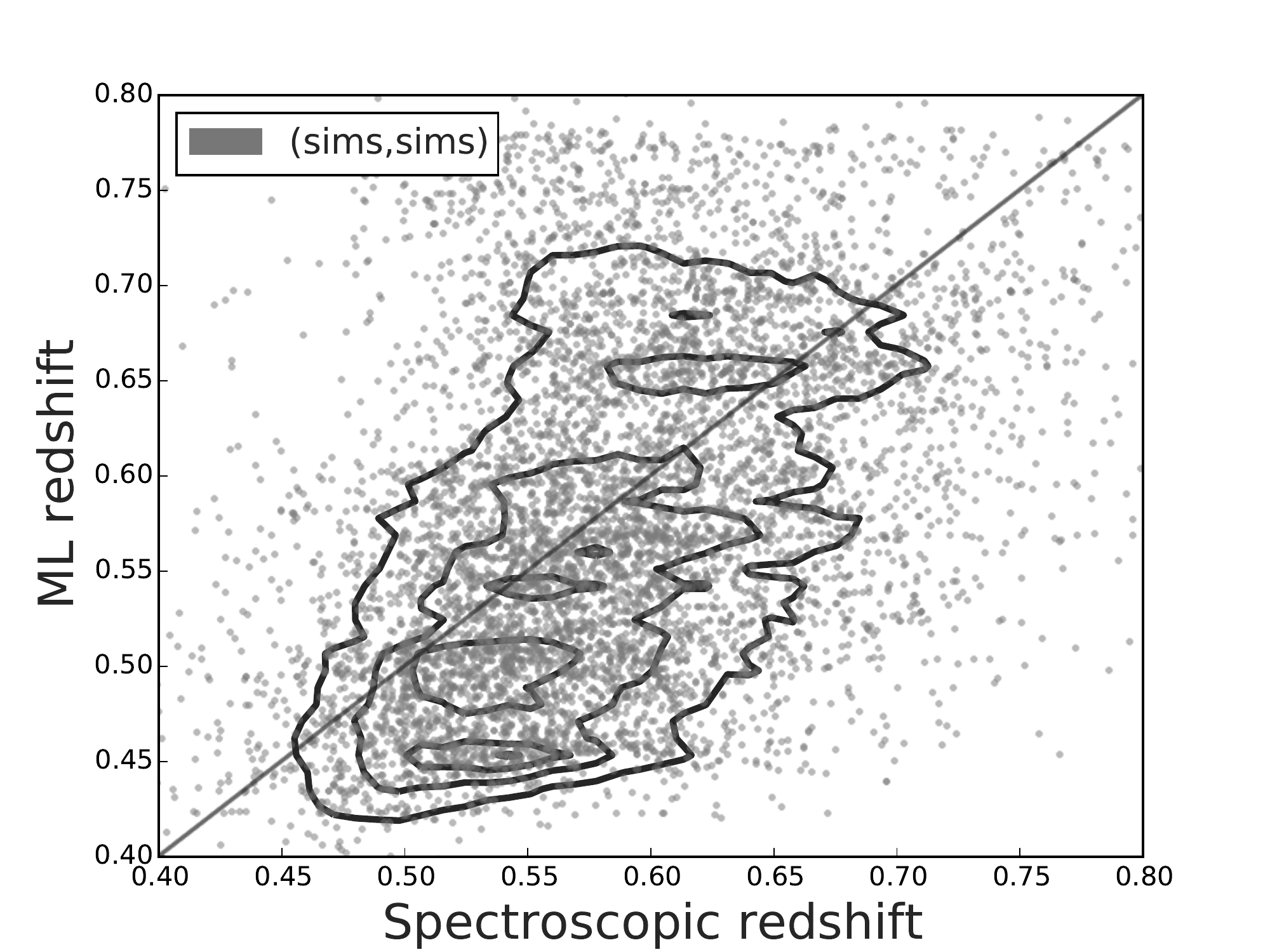}\\
 \includegraphics[scale=0.475, clip=true, trim=5 00 50 35]{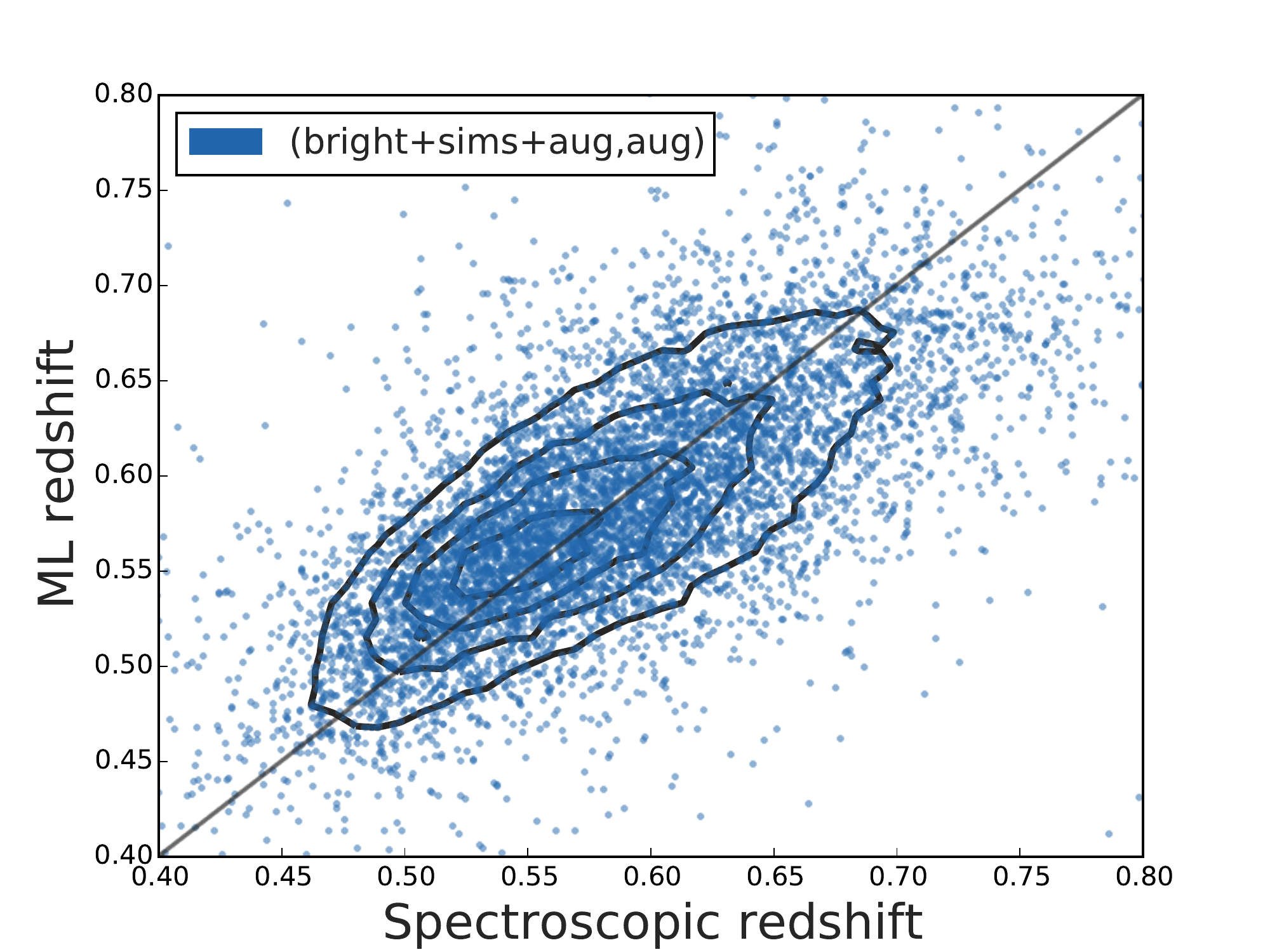}
 \includegraphics[scale=0.475, clip=true, trim=70 00 50 35]{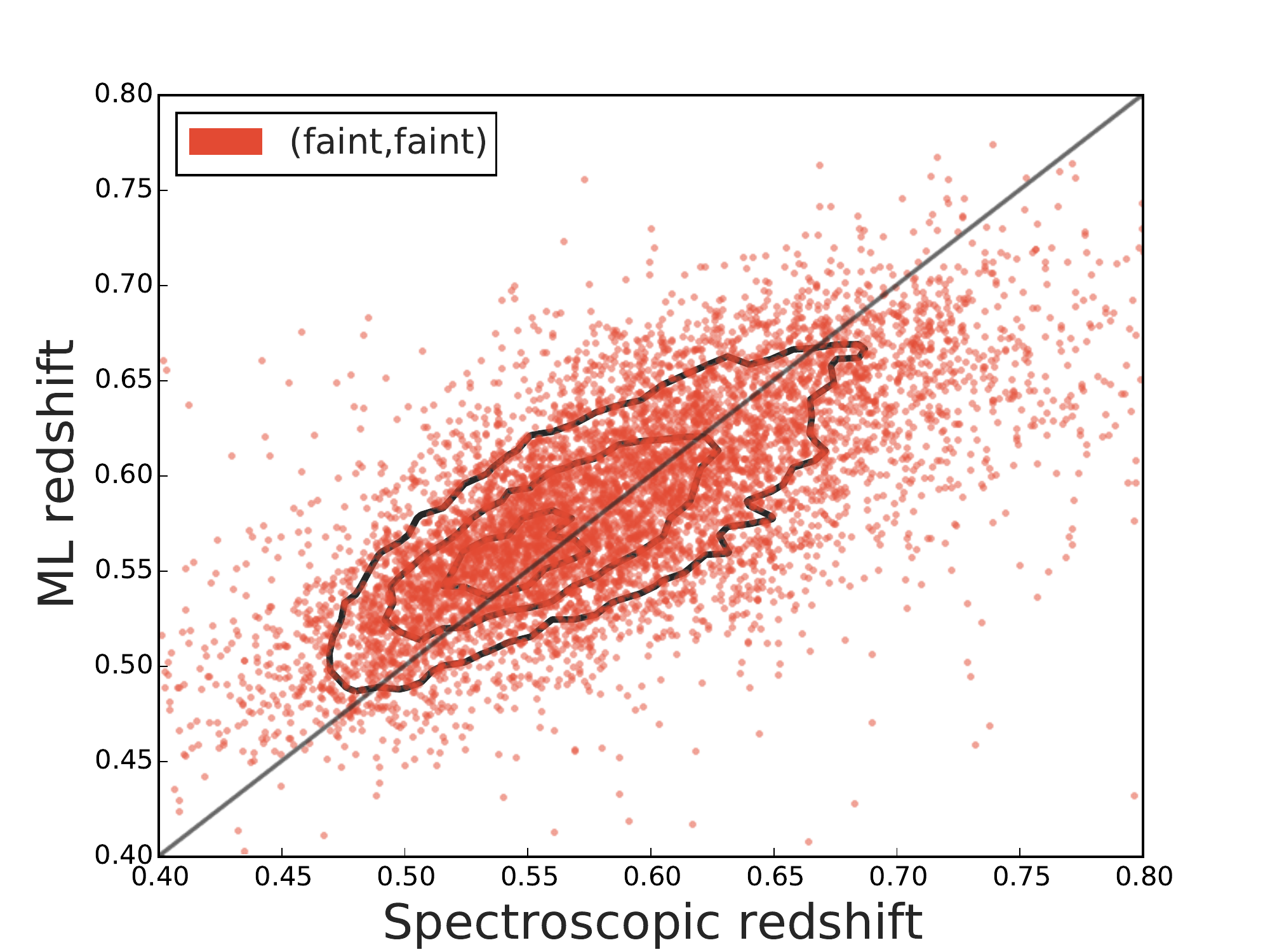}
  \caption{ \label{Ml-spec} The machine learning redshift of the faint sample of SDSS DR10 test galaxies against the spectroscopic redshift, for different combinations of training  and cross-validation samples as shown in the legend of each panel. See Fig. \ref{redshiftPlots} for keyword definitions. The test galaxies are not used during training, and represent the ability of the learned system to estimate redshifts for the test galaxies which fall outside of the original (bright) training set.}
\end{figure*}

In Fig. \ref{Ml-spec} we show the machine learning redshift of the faint sample of test SDSS DR10 galaxies against the spectroscopic redshift, for different combinations of training and cross-validation samples, as shown in the legend of each panel. We find that the scatter is the smallest in the ideal case (faint, faint) of training and cross-validating on the faint galaxies, the simulations has the largest scatter, and the augmented data case has a scatter between the no augmentation case (bright, bright) and the ideal case. We show the redshift range $0.4<z<0.8$ which contains most of the faint galaxies.

Finally we compare these results obtained using data augmentation applied to the DR8 (DR10) faint samples, with the  photometric redshifts available from within SDSS CasJobs \citep[][using a hybrid k-NN \& template approach]{2000AJ....120.1588B,2007AN....328..852C,2009ApJS..182..543A} for the same galaxies. Using the SDSS photometric redshifts we find the values $\sigma_{68} =$ 0.026 (0.037)  and an outlier rate of 2.30\% (2.70\%) for the DR8 (DR10) analyses.  For DR10 the measured values of $\sigma_{68}$ are improved using data augmentation but for DR8 the values of $\sigma_{68}$ are similar. In both DR8 and DR10 we find that the outlier fraction is reduced using the data augmentation procedure and forests. Therefore we conclude that, remarkably, we find that assuming no knowledge of real galaxies, but using data augmentation, actually improves the redshift estimates compared with the standard SDSS machine learning photometric redshifts which does train on real galaxies. These results are probably more due to the machine learning architecture used, than the data augmentation process, see e.g. Fig. 4 of \cite{2015MNRAS.449.1275H}.

\begin{figure*}
 \includegraphics[scale=0.475,clip=true,trim=5 40 50 35 ]{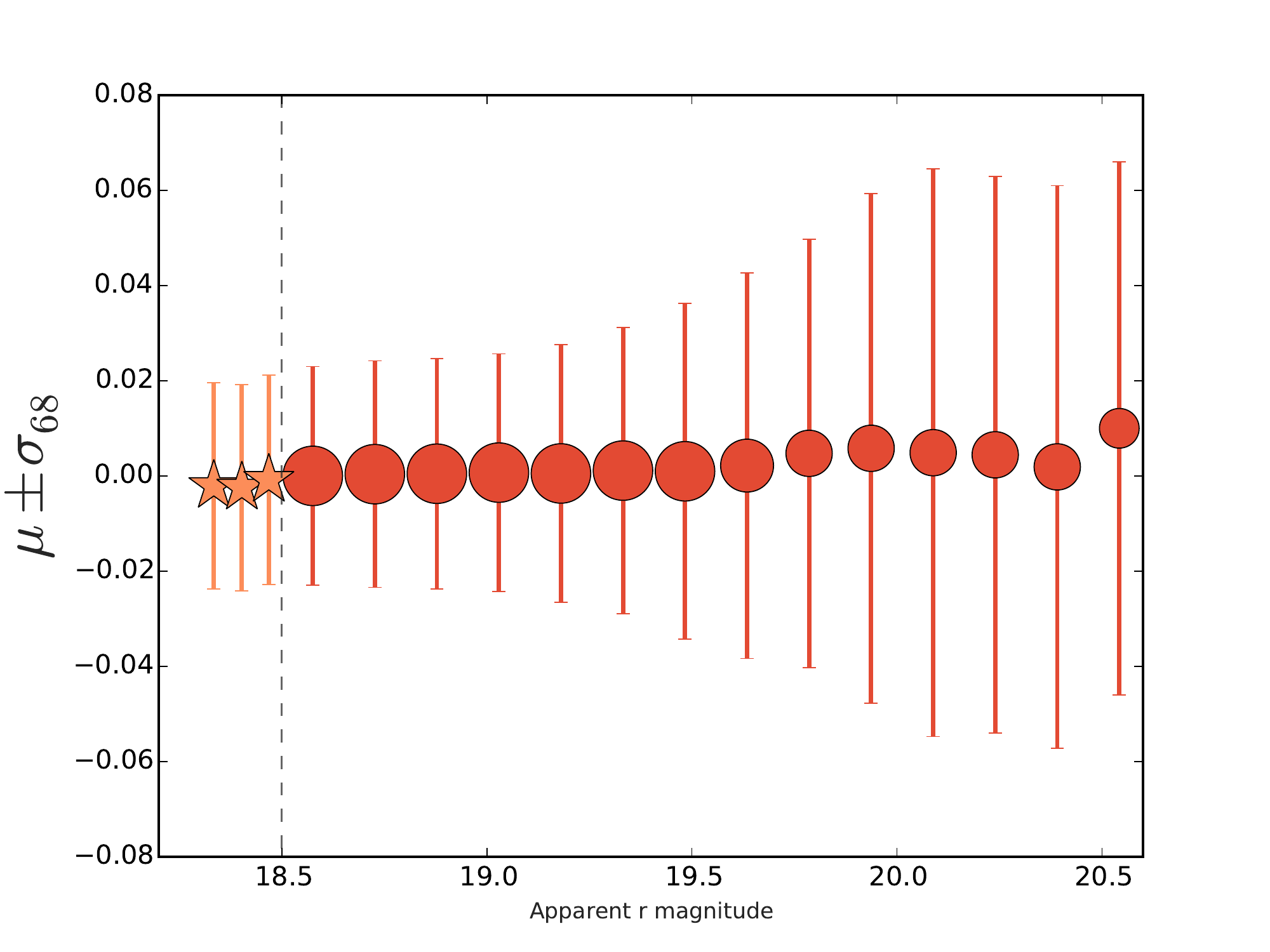}
  \includegraphics[scale=0.475,clip=true,trim=5 40 42 35 ]{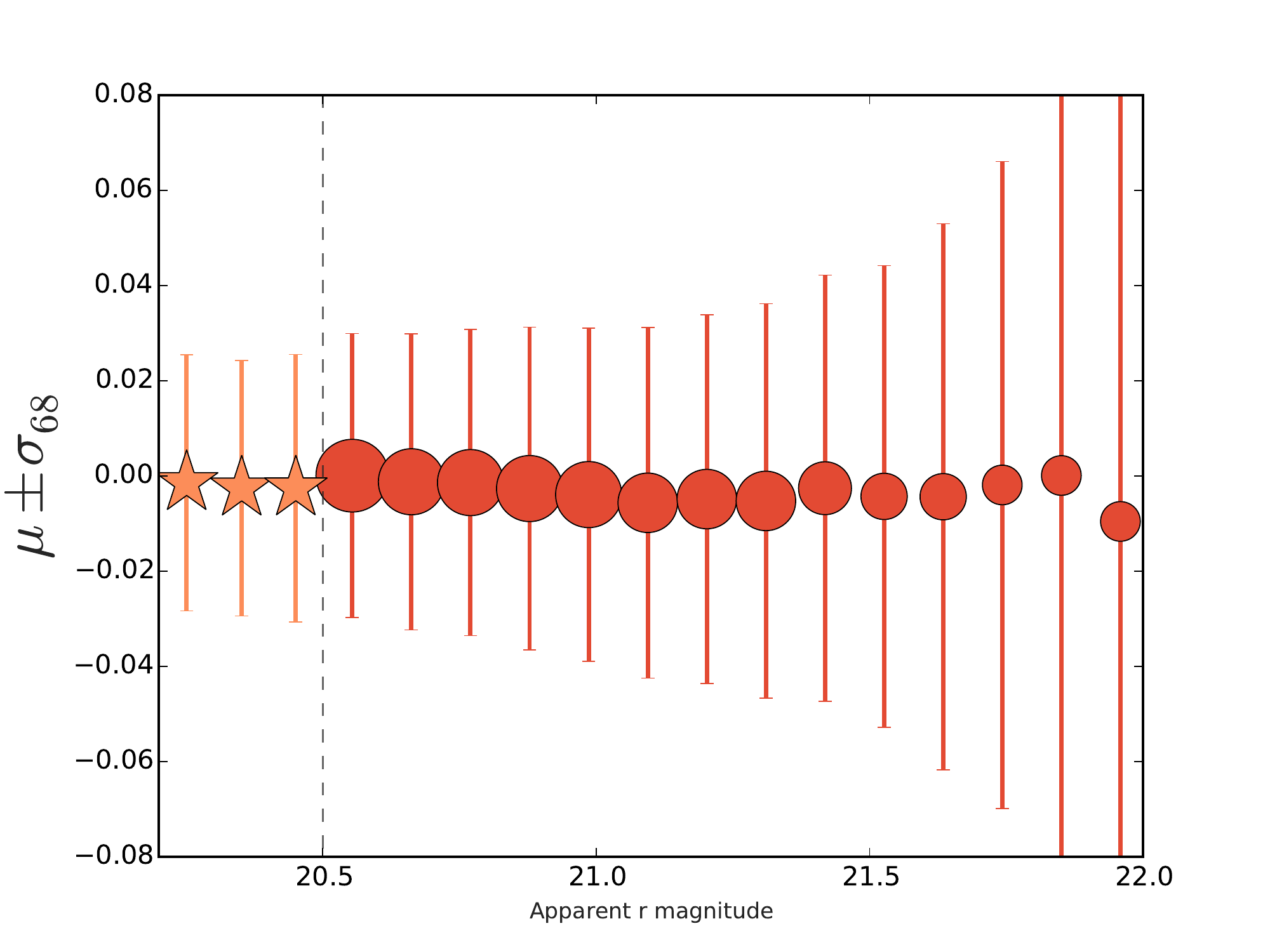}\\

 \includegraphics[scale=0.475, clip=true, trim=5 0 50 42]{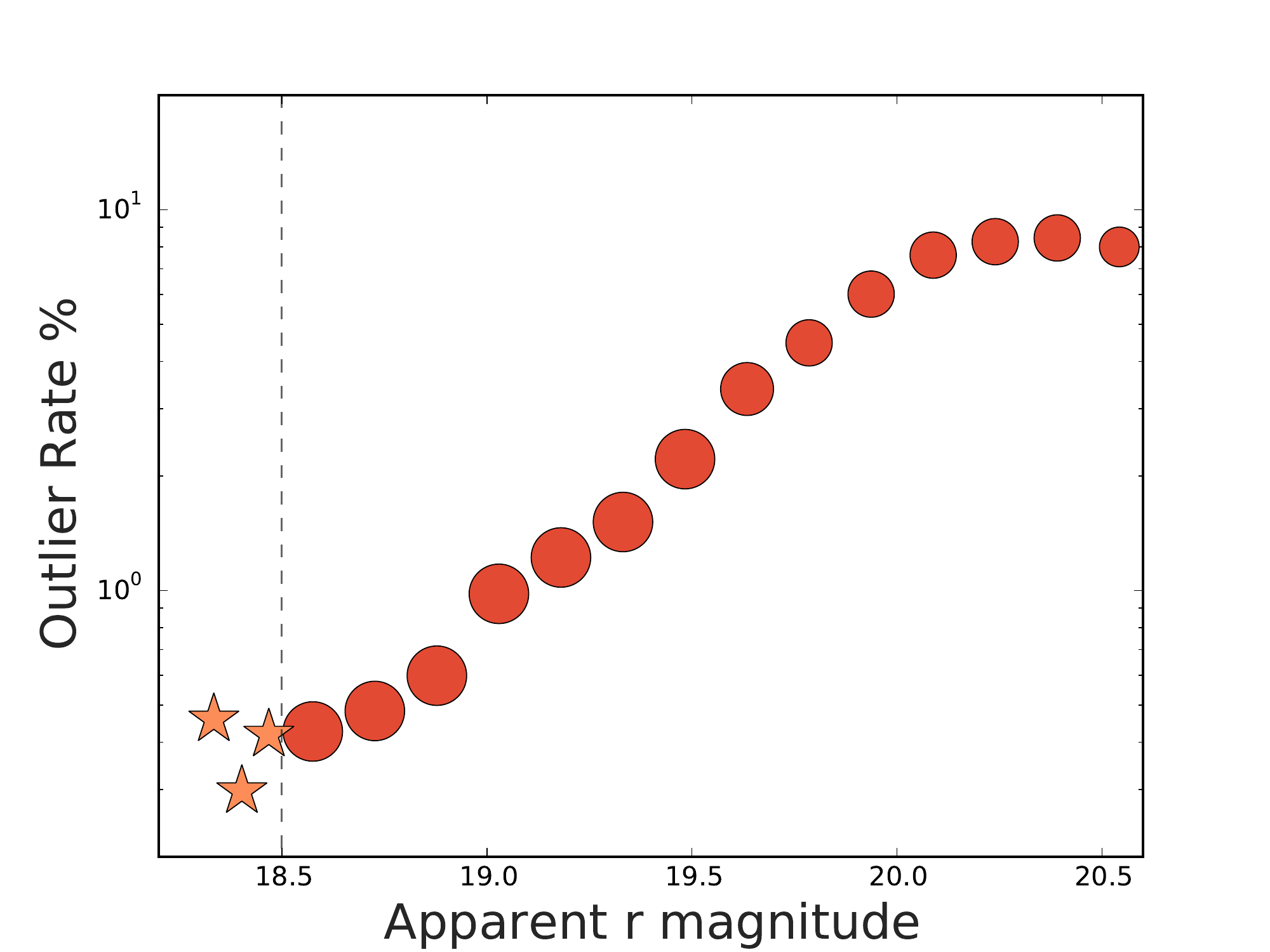}
 \includegraphics[scale=0.475, clip=true, trim=5 0 42 42]{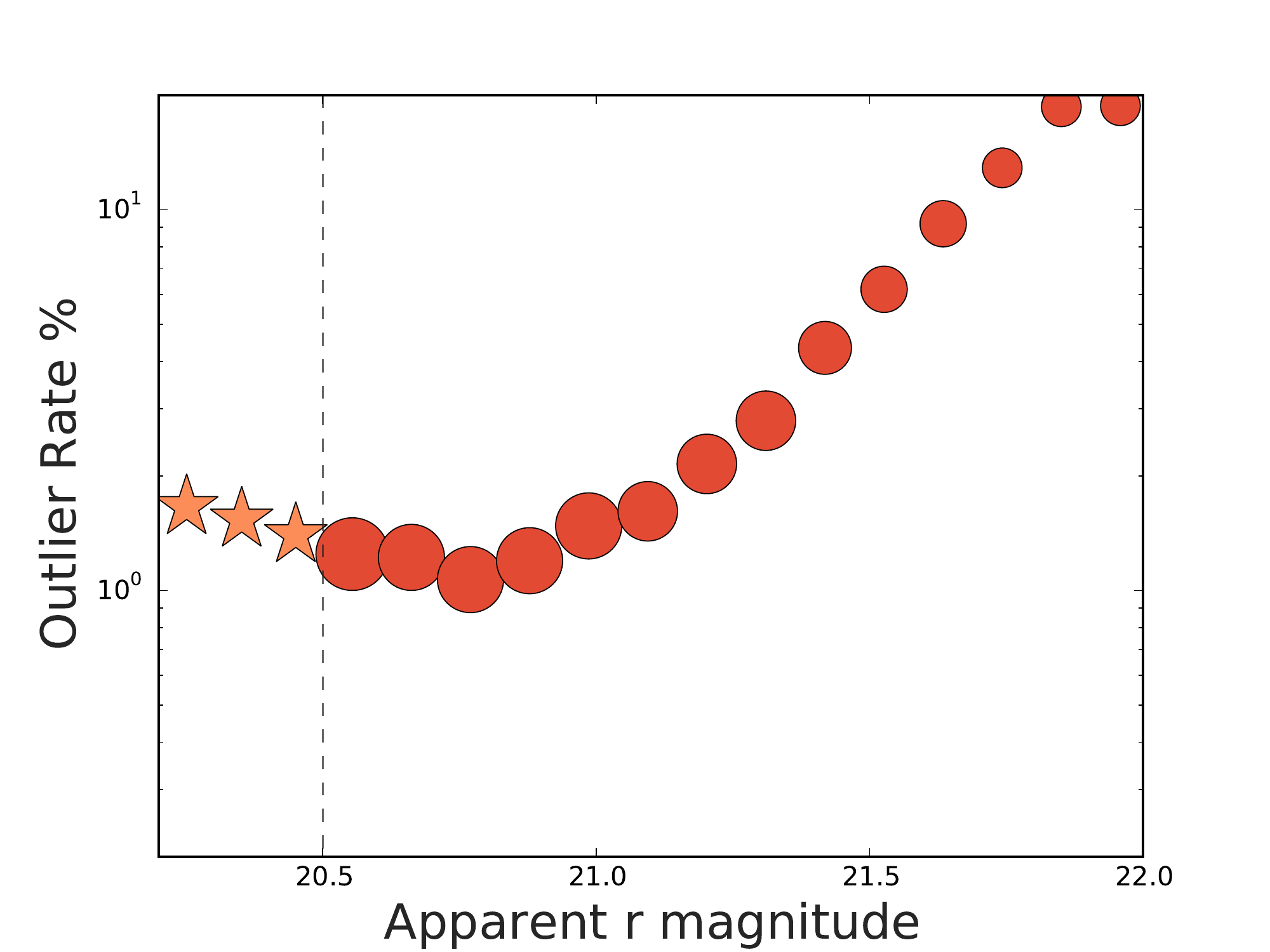}\\
 \caption{ \label{r-dep} The effect on the recovered redshifts as measured on the faint galaxy sample as we probe to increasingly deeper $r$ band magnitudes, past the artificially imposed magnitude limit of training sample (dashed line). The left (right) panels show the SDSS DR8 (DR10) analysis. Top panel shows the median value $\mu$, and the dispersion measured by $\sigma_{68}$ of the redshift scaled residual distributions $\Delta_{z'}$. The lower panels show the outlier rate defined as $|\Delta_{z'}|>0.15$. The starred data points to the left of the vertical dashed line are measured from the bright data, without data augmentation. The data points to the right of the vertical dashed line are measured from the faint data using data augmentation. The area of the symbol is proportional to the square root of the number of test set galaxies in the magnitude bin. }
\end{figure*}
{\rr
\begin{figure*}
 \includegraphics[scale=0.475,clip=true,trim=5 40 50 35 ]{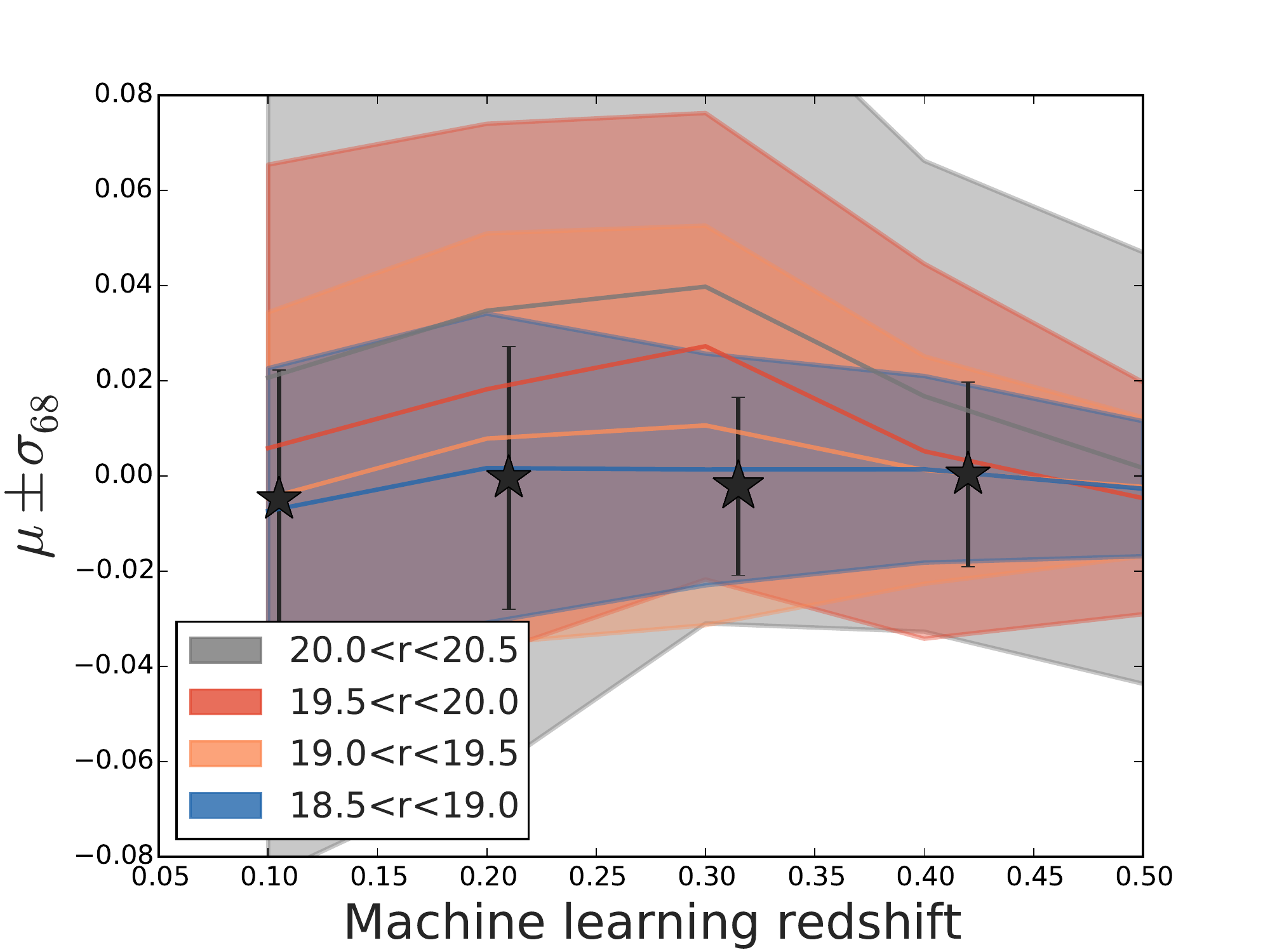}
  \includegraphics[scale=0.475,clip=true,trim=5 40 42 35 ]{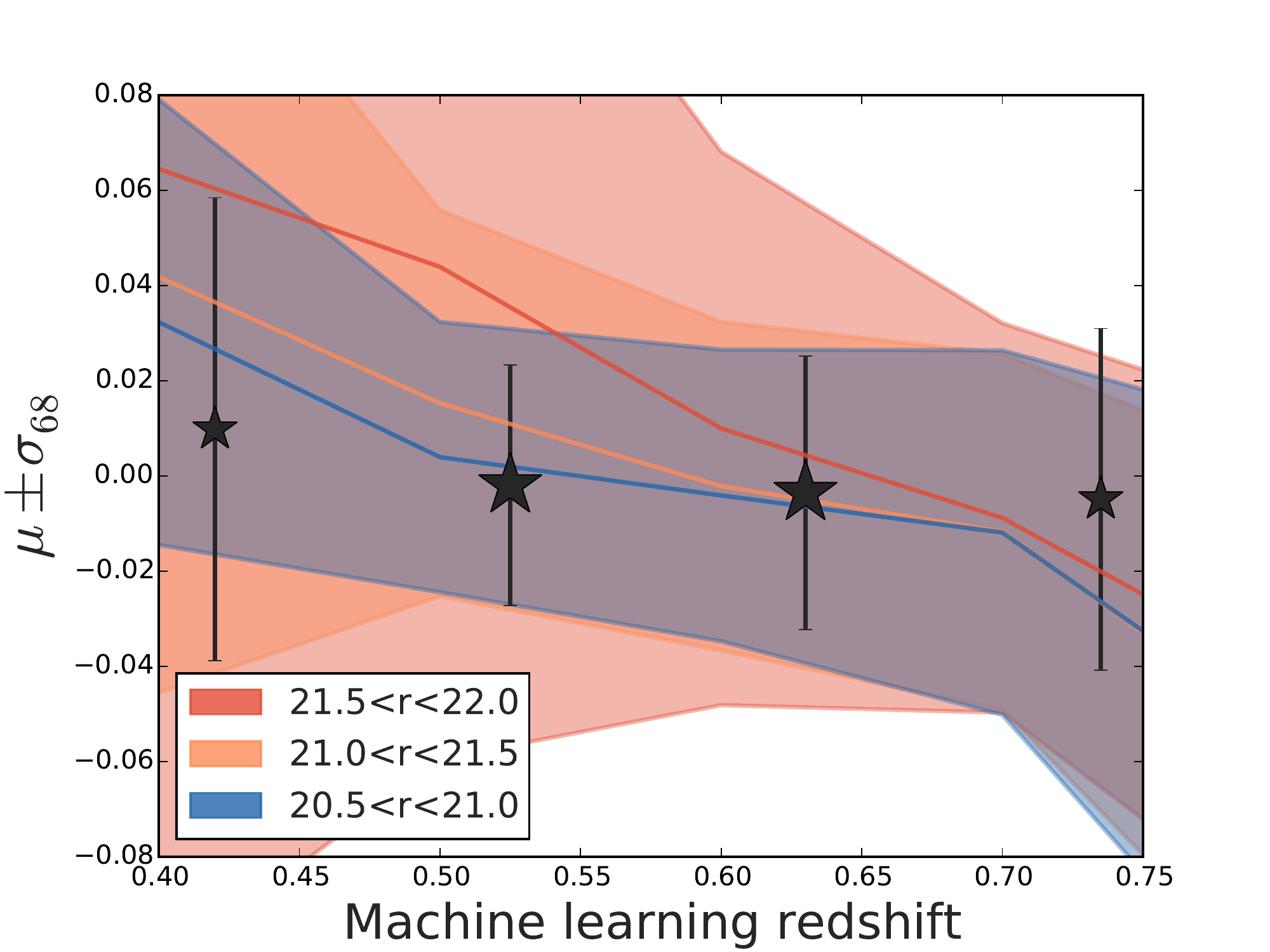}\\
 \includegraphics[scale=0.475, clip=true, trim=5 0 50 42]{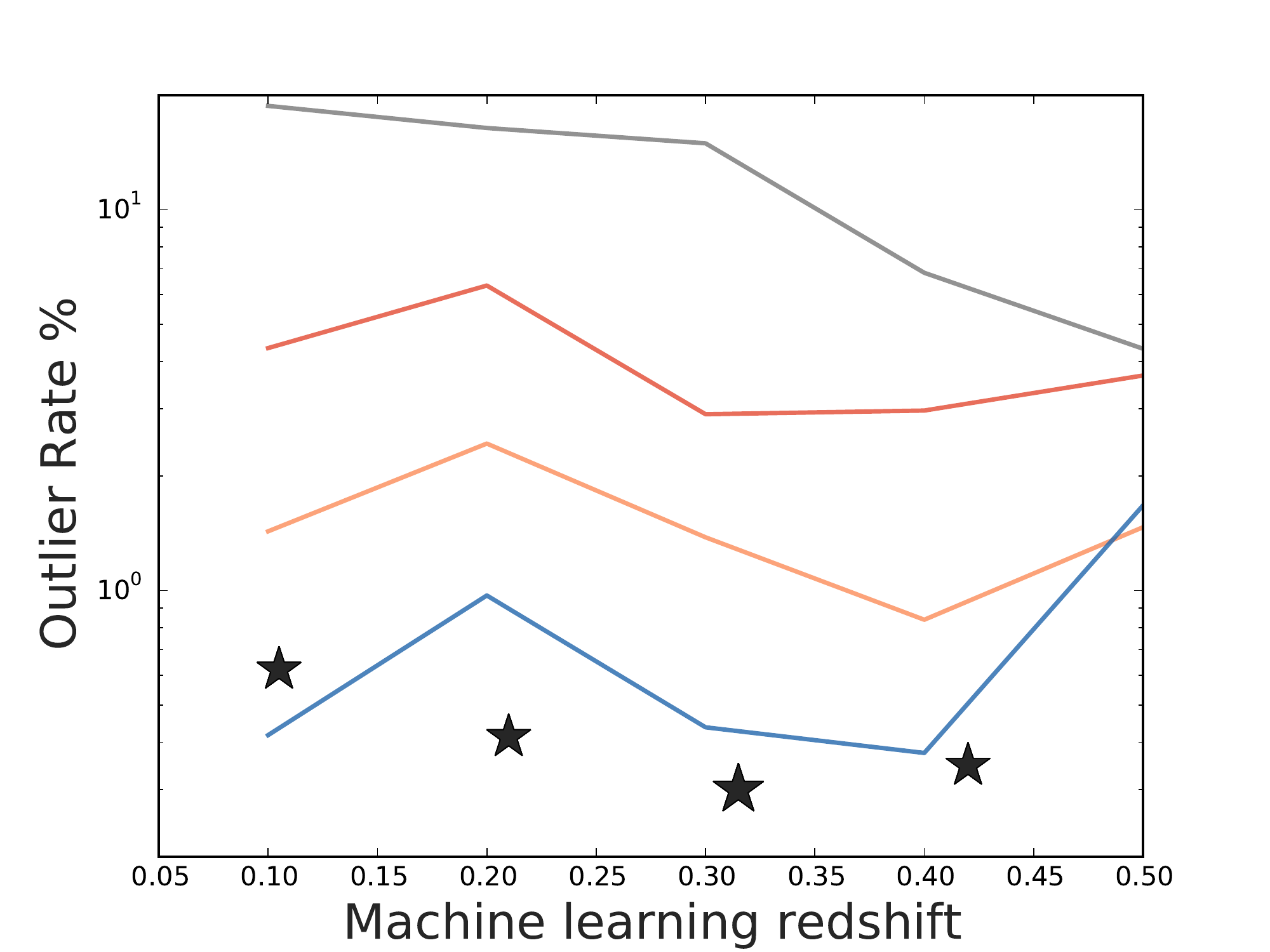}
 \includegraphics[scale=0.475, clip=true, trim=5 0 42 42]{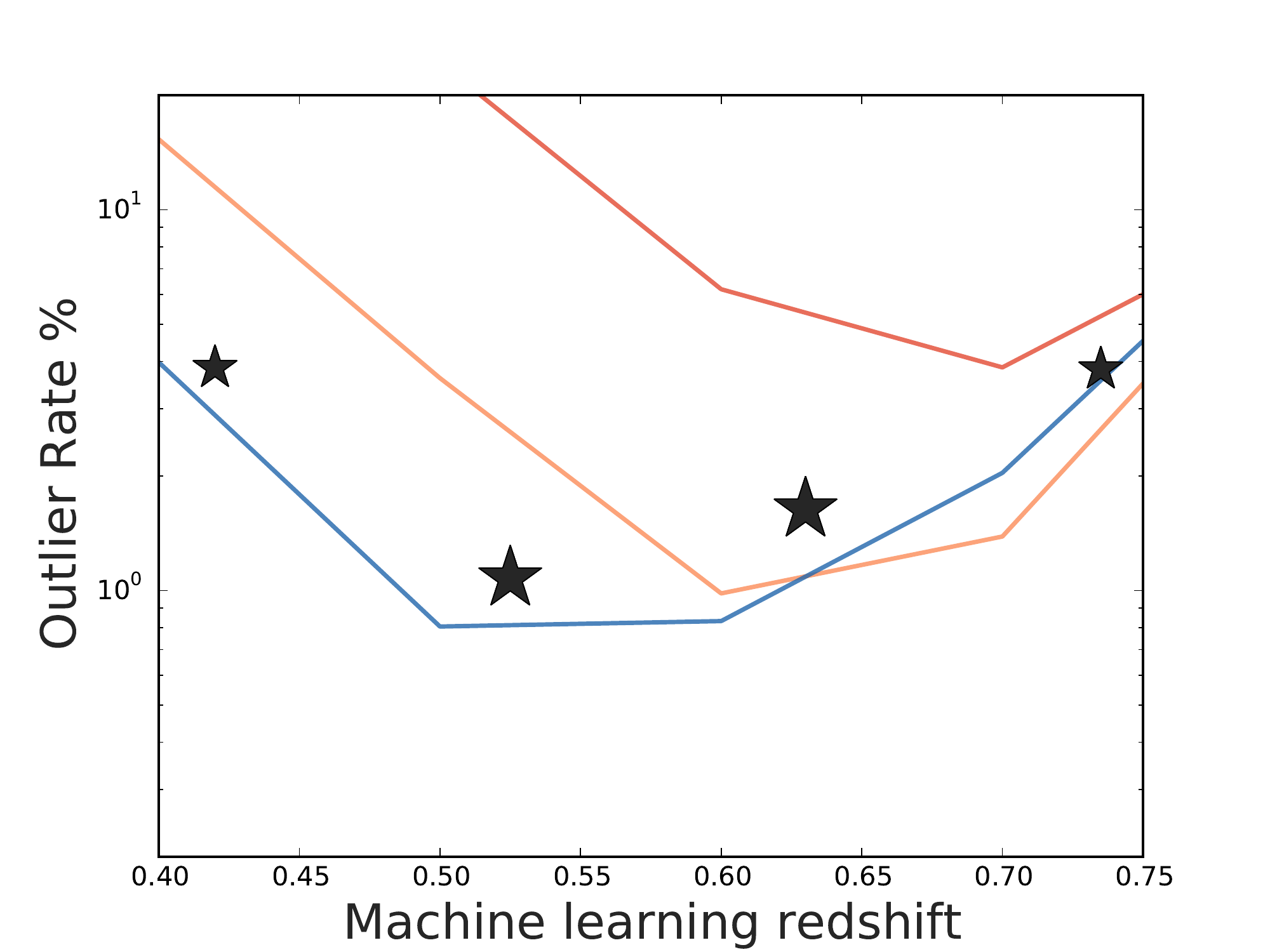}\\
 \caption{ \label{zr-dep} {\rr The effect on the recovered redshifts as measured on the faint galaxy sample as a function of machine learning redshift and the apparent $r$ band magnitude. The starred data points show the results of the same benchmark sample as Fig. \ref{r-dep} subdivided by redshift not apparent magnitude and are measured from the bright data, without data augmentation. The left (right) panels show the SDSS DR8 (DR10) analysis. Top panels show the median value $\mu$, and the shaded region shows the dispersion measured by $\sigma_{68}$ of the redshift scaled residual distributions $\Delta_{z'}$. The lower panels show the outlier rate defined as $|\Delta_{z'}|>0.15$. } }
\end{figure*}
}

\subsubsection{Nearest Neighbor selection of augmented data}
\label{knnsearch}
We further explore the effect of more carefully selecting all of the augmented data sets such that their input features (i.e., the magnitudes and colors) are similar to the faint test sample. The samples are chosen by selecting the three nearest neighbors in each of the training and cross-validation augmented data sets, to each of the faint galaxies in the test set. The k-NN nearest neighbor search algorithm \citep[see e.g.][]{doi:10.1080/00031305.1992.10475879} is used to perform the matching. 
The above analysis is then repeated and the results are analysed. We find that this method of carefully selecting the training and cross-validation samples does not noticeably affect the final results.

\subsection{The effect of probing to deeper magnitudes}
\label{deepermags}
We next explore how the recovered redshifts degrade as we probe to increasingly deeper $r$ band magnitudes, past the artificially imposed magnitude limit of the training sample. This is analogous to extrapolating deeper into the unpopulated $r$ band magnitude and corresponding color dimensions, while relying more heavily on the data augmentation to provide estimates for how galaxies in these parts of input feature space scale with redshift.

For this section we again perform two independent analyses using SDSS DR8 and  DR10 data. However we now use the entire faint galaxy sample as the test sample. This corresponds to 124k galaxies for SDSS DR8 and 300k galaxies for SDSS DR10. We can use the full test sample because the faint samples are not used during the training and cross-validation, and we will not compare results with the ideal case. In both sets of analyses we pass the faint test data through the best forest found in the previous section determined using the bright data, augmented data, and simulations as a training sample, and the augmented data as a cross-validation sample. 

We group the test data into bins of apparent magnitude $r$, and construct $\Delta_{z'}(r)$ by comparing the machine learning redshift of the galaxies in each bin with the spectroscopic redshift. We measure $\mu,\sigma_{68}$ and the outlier rates from  $\Delta_{z'}(r)$, and present the results in Fig. \ref{r-dep}. The SDSS DR8 analysis is again shown in the left panels, and the DR10 analysis in the right panels. The area of the plotting symbol is proportional to the square root of the number of data in each magnitude bin.

To make the results of these analyses more applicable to other datasets, we construct benchmark samples from the bright galaxy datasets. The benchmark sample correspond to bright galaxies near the artificially imposed boundary. For the DR8 analysis we construct a benchmark sample from the bright galaxies within the magnitude range $18.2<r<18.5$ which consists of 22.5k galaxies, and for the DR10 analysis the benchmark is constructed from galaxies within the magnitude range $20.3<r<20.5$ and consists of 175k galaxies. We construct random samples of training, cross-validation and test samples of size (50\%, 25\% and 25\%). We perform a standard machine learning redshift analysis on the benchmark samples, similar to that described in \S\ref{augemnet_all_data} but without using the data augmentation. The results of the benchmark analysis are shown in the panels of Fig. \ref{r-dep} and correspond to the starred data points to the left of the vertical dashed line, which marks the divide between the bright and faint galaxy samples. 

The top panels in Fig. \ref{r-dep} show how the values of the median $\mu$ and $\sigma_{68}$ change as we probe to deeper $r$ band magnitudes past the magnitude limit of the bright sample. As we might expect, we find that the recovered values of $\sigma_{68}$ degrade the deeper we probe past the magnitude limit of the training sample. We find that using data augmentation results in a {\rr well controlled, small valued bias} for the redshift estimates even as we probe more than 1.5 magnitudes deeper past the artificial magnitude limit. This is true for both sets of analyses. 

The bottom panels of Fig. \ref{r-dep} show that the outlier rates also increase as we move to deeper magnitudes. This effect is again seen in both sets of analyses. Comparing the starred points of the benchmark sample to the left of the dashed line with the data augmentation analyses to the right of the dashed line, we find that the outlier rate for the DR8 (DR10) analysis is very similar directly across the magnitude limit, but steadily degrade to within a factor of 5 (2) at a depth of one magnitude past the limit. 

We remark on the success of data augmentation to estimate redshifts of galaxies which are fainter than the original training set. We find that for all magnitude bins examined here the size of $\sigma_{68}$ is within a factor of $\approx2$ of the boundary data set, when one attempts to measure galaxy redshifts one magnitude deeper than the limit of the training sample. The errors degrade further if one attempts to recover redshift estimates 1.5 magnitudes and 2 magnitudes deeper than the training and cross-validation samples. These results suggest that we can use data augmentation to explore past the magnitude limit of a training galaxy sample, if we are willing to accept a degradation in the size of the recovered redshift error. However we are likely to produce redshift estimates with {\rr low bias}.

{\rr 
\subsection{The effect of probing to deeper magnitudes as a function of redshift}
We next examine the effect of the recovered machine learning redshift as a function of both apparent $r$ band magnitude and the estimated machine learning redshift, $z$. We calculate $\Delta_{z'}(z,r)$ in redshift slices of width 0.1, and apparent magnitude bins of width 0.5 past the artificially imposed magnitude limit, and show these results in Fig. \ref{zr-dep}. The chosen magnitude bin ranges are shown in the legend, and the left panels show the results for the SDSS DR8 analysis and the right panels show the results for SDSS DR10. We again calculate benchmark values of $\Delta_{z'}(z,r)$ using bright data which is just brighter than the apparent magnitude limit, see \S\ref{deepermags}, but now subdivided this sample by machine learning redshift not apparent $r$ band magnitude.  We show these benchmark values as a function of redshift using the starred data points in Fig. \ref{zr-dep}.

Fig. \ref{zr-dep} indicates that the measured statistics at each redshift of the benchmark sample, most closely resemble those of the faint galaxy sample which are closest to the artificial limiting magnitude. In particular the data augmentation applied to galaxies with a $r$ band magnitude up to 0.5 magnitudes deeper than the benchmark sample, is very well controlled. The median, spread of the data $\sigma_{68}$, and outlier fraction of $\Delta_{z'}$, differ little from the starred data points. Examining the results of data in the magnitude bin $0.5<r<1.0$ magnitudes deeper than the benchmark sample, we find that all measured statistics of $\Delta_{z'}$ degrade. The outlier fraction of this sample is increased by a factor of a few for the SDSS DR10 analysis to a factor of 10 for the SDSS DR8 analysis.

Using data augmentation to extrapolate to deeper apparent magnitudes shows that both the errors and bias increase and the outlier fraction increases by more than an order of magnitude compared to the bench mark sample.
}
\section{Discussion}
\label{diss}
We next discuss and interpret these results in terms of photometric depth and SED templates, and with respect to other works.

Examining Fig. \ref{Ml-spec} we find that using K-correct templates to augment the bright data improves the redshift estimates of the faint test sample. This suggests that the galaxy population which we create using K-correct is a reasonable approximation to that of the true underlying galaxy population. This also implies that the change in the galaxy populations are smooth arcoss the redshift ranges and apparent magnitude depths explored.  A similar result is described in \cite{2013MNRAS.432.1046B} (see their Fig. 3) using shallow CFHTLS-Wide photometric data, together with deep photometric CFHTLS-D data and spectroscopic samples. The templates were chosen to optimally match photometric redshift estimates to spectroscopic ones for CFHTLS-W galaxies with magnitudes $i<22$. If the same templates were then applied to a fainter CFHTLS-W sample with $22<i<24$, the redshift accuracy deteriorated. However when (for overlapping fields) the CFHTLS-W photometry was replaced with that of the deeper CFHTLS-D survey, the redshift accuracy improved and reached the same value as for the bright, $i<22$ CFHTLS-W sample. The authors conclude that the limiting
factor in measuring photometric redshifts for faint galaxies is the signal to noise (or depth) of the photometry and not the template set used.

The peak of the redshift scaled residual histogram (Fig. \ref{redshiftPlots}) for the simulations is slightly offset from the line $x=0$ and from the other distributions. One can think of the simulations as being a realization of the template methods for galaxies at different redshifts, and with a range of physical properties. The semi-analytic models use SED templates which encode stellar physics and our knowledge about galaxy evolution to produce realizations of these models as simulated galaxies. This slight redshift offset noted above has been seen by others when estimating photometric redshifts using templates methods applied to SDSS galaxies \citep[][]{2013ApJ...768..117G}.

{\rr We note that semi analytic models are not observed data: they use models to extrapolate observed galaxy properties between redshifts, and therefore may not encode real observational effects e.g., from the evolution of multi-band colors and number densities. Another reason for a color mismatch in the semi-analytic models can arise if recipes which relate the star formation and feedback processes to merging of halos are not correct. Furthermore the simple stellar population models often assume a single epoch of star formation. However the fact that data augmentation using simulations provides reasonable redshift estimates suggest that the current models do provide a good approximation to these effects.
 }

Finally we note that none of the data augmentation methods is better than using nature itself; the (faint, faint) ideal case produces the best of the machine learning redshift estimates examined here. This reaffirms that we can still improve our stellar populations models, and templates in order to more closely mimic galaxies observed in nature \citep[see also][]{2011MNRAS.413..434T,2013ApJ...768..117G}.

%% file: conclus.tex
Photometric surveys can be maximally exploited for large scale structure analysis once galaxies have been identified and their positions on the sky and in redshift space measured. Very accurate spectroscopic redshifts are only measured on a small and often biased subset of the full photometric galaxy sample due to the integration times required to obtain a reliable measurement, and the spectroscopic magnitude limit of the survey typically being shallower than the photometric detection magnitude limit.

This implies that if one attempts to estimate photometric redshifts of all galaxies using a machine learning architecture, one may be applying the results of a spectroscopic training sample which is not fully representative of the final photometric test sample. For the Sloan Digital Sky Survey \citep[SDSS, ][]{2000AJ....120.1579Y} this corresponds to a limit in the apparent $r$ band magnitude.

An alternative to machine learning methods is template methods, which can also estimate the photometric redshift of galaxies of all magnitudes, including those which are deeper than a spectroscopic training sample, however such a sample need not even exist. The templates encode our physical knowledge of galaxy color evolution through stellar population models. 

Remaining within the machine learning framework, one may use the knowledge gained from stellar population models to simulate galaxy properties using semi-analytic models, and augment (or compliment) the original spectroscopic training sample using the simulated galaxies. The reason to augment the training and cross-validation data is to ensure that they populate the same input feature parameter space as the final test sample. Even though this augmentation is a `best guess' of how the galaxies should appear, the process uses testable physics (if the templates are based on stellar population models), and it is still preferable to not having any training or cross-validation data in the input feature parameter space occupied by the test data. 

Another approach to augment observed data is by using the public SDSS K-correct package \citep[][]{2007AJ....133..734B}. One may even use spectra obtained by other sources, and estimate their magnitudes as if they were to have been observed in the photometric survey in question. This last approach is being actively explored by the authors, \cite[see also][]{2004A&A...423..761V}. {\rr If only the feature space number density of training galaxies is biased compared to the test galaxies, one may use galaxy re-weighting schemes \citep[e.g.,][]{2008MNRAS.390..118L,2009MNRAS.396.2379C} or the covariate shift method \citep[][]{2014arXiv1406.4407S}.}

In this paper we present a study of the effect on the recovered machine learning redshift applied to a non-representative sample of test galaxies which are selected to be fainter in the $r$ band than the training sample of galaxies. We perform two sets of analyses, drawing on 800k galaxies from the SDSS DR8 and 1.7 million galaxies from SDSS DR10. The main difference between these data samples are that the DR10 sample probes higher redshifts and deeper $r$ band magnitudes. We apply a $r$ band apparent magnitude limit of 18.5 (20.5) for the DR8 (DR10) galaxies to identify bright training and bright cross-validation samples, and faint test samples.

We augment the bright galaxy training and cross-validation data in both sets of  analyses using the latest semi-analytic models \citep[][]{2014arXiv1410.0365H} which separately incorporate two different stellar populations models \citep[][]{2003MNRAS.344.1000B,Maraston:2004em} applied to the Millennium Simulation \citep[][]{2005Natur.435..629S}. We further augmented the bright galaxy training and cross-validation samples using the SDSS K-correct package, which estimates the apparent magnitude of an observed galaxy if it were to exist at a different redshift. 

By choosing to perform data augmentation with existing simulations and with K-correct we are restricted to the available input features. These features are magnitudes and colors. Recent work has shown that the use of additional features can lead to improvements in the recovered redshift estimates \citep{2015MNRAS.449.1275H}. It would be interesting to determine if the construction of additional  input features such as radii and galaxy shapes could also lead to a further improvement in machine learning redshift estimates through the use of data augmentation.

We explore combinations of training and cross-validation samples, for example by using the bright data, K-correct augmented data, and simulated data as a training sample, and the K-correct augmented data as a cross-validation sample. For each combination we tune the hyper-parameters of the machine learning architecture. Finally we pass the faint test data into the machine to estimate a machine learning photometric redshift $z$ and calculate $\Delta_{z'}=(z-z_{spec})/(1+z_{spec})$. We determine the value $\sigma_{68}$ which contains 68\% of the data about the median value $\mu$ of $\Delta_{z'}$, and also calculate the outlier rate defined as the fraction of data with $|\Delta_{z'}|>0.15$.

The machine learning architecture is chosen to be decision tress for regression, of which many trees are grown with the algorithm Adaboost \citep[][]{ig,Freund1997119,Drucker:1997:IRU:645526.657132} which we refer to as a forest.  We define two benchmark samples corresponding to the worst case (no data augmentation) and the ideal case (the training, cross-validation and test sample are all drawn from the faint sample). We present our results with respect to these two benchmark cases.

We find that the use of the augmented data sets improves the error $\sigma_{68}$, on machine learning redshift estimates by 41\% in both DR8 \& DR10 sets of analyses, when compared to the ideal case.  This means that using data augmentation we are able to improve the redshift estimates from the worst case (bright samples) and recover up to 41\% of the possible improvement that we may hope to achieve if we had the ideal case (faint samples).

We find the outlier rate is also improved by 80\% for the SDSS DR8 analysis and 10\% for the SDSS DR10 analysis. It is satisfying to note that using only the simulated galaxies as training and cross-validation samples we still recover a reasonable photometric error of $\approx 1.7 \times \sigma^{ideal}_{68}$, and an outlier rate of only a factor of 2 (4) higher than the ideal case when applied to the SDSS DR8 (DR10) faint test sample. This shows how accurately semi-analytic models can replicate the magnitudes, colors and redshifts of observed galaxies.

We also compare these results with the photometric redshifts available from within SDSS CasJobs \citep[][]{2000AJ....120.1588B,2007AN....328..852C,2009ApJS..182..543A} for the same galaxies. The SDSS photometric redshifts are trained on real galaxies, while data augmentation trains only on augmented galaxies. For DR10 the measured values of $\sigma_{68}$ are improved by 10\% using data augmentation. However for DR8 the values of $\sigma_{68}$ remain very similar. In both DR8 and DR10 we find that the outlier fraction is reduced by $\approx30\%$ using the data augmentation procedure and forests. 

We next explore the change in the recovered redshift errors and outlier rate, as we probe to increasingly deeper $r$ band magnitudes, past the artificially imposed magnitude limit of the bright training samples. This makes the training and test samples more dissimilar. We find that as one pushes deeper past the $r$ band magnitude limit, both the error $\sigma_{68}$ and the outlier rate slowly degrade. We note that the median values of the $\Delta_{z'}$ are always close to 0, and therefore the combination of data augmentation and tree based methods produce only a {\rr small value of bias} in the estimate of the true galaxy redshift, even when applied to an unrepresentative test sample of galaxies. {\rr We do however note that the level of bias reported here (using the SDSS data set) is larger than the science requirements of the Dark Energy Survey \citep[][]{2005astro.ph.10346T}. Further investigation is underway by the authors to understand if data augmentation can achieve the specified level of precision.

Applying these analyses to galaxies with an $r$ band magnitude up to 0.5 magnitudes deeper than the artificially imposed magnitude limit of the benchmark training samples result in very well controlled statistics for all machine learning redshifts ranges explored. The median, spread of the data $\sigma_{68}$, and outlier fraction of $\Delta_{z'}$, differ little from the benchmark samples. Examining the results of data in the magnitude bin $0.5<r<1.0$ magnitudes deeper than the benchmark sample, we find that all measured statistics of $\Delta_{z'}$ degrade. The outlier fraction of this sample is increased by a factor of a few for the SDSS DR10 analysis to a factor of 10 for the SDSS DR8 analysis.
}

Although one should not extrapolate these results to new or different data sets, we do expect data augmentation to also improve other similar analysis.  In this paper we have concentrated on the improvement to the point estimate of the redshift while using data augmentation. The full shape of the redshift distribution function, or the conditional probability distribution function for individual galaxies are also important quantities which we will examine in future work \citep[][]{RauEtAllinPrep}. 

Finally we conclude that the use of data augmentation presents an accessible way to include our knowledge of the Universe, and in particular the magnitude evolution of galaxies, into the machine learning training and cross-validation samples. This is particularly important if the training and cross-validations samples are drawn from non-representative samples of the final test data. We have shown that this method works well for SDSS galaxies using two sets of analyses. Similar analysis could be performed on other data sets and surveys, in particular for surveys with existing dedicated simulations.